\documentclass[pra,twocolumn,showpacs,preprintnumbers,superscriptaddress,
amsmath,amssymb,tightenlines,epsfig]{revtex4}

\usepackage{bm}
\usepackage{graphicx}

\renewcommand{\i}{\int}

\renewcommand{\d}{\dag}
\newcommand{\h}{\hat}
\newcommand{\p}{\partial}
\renewcommand{\v}{\vert}

\newcommand{\f}{\frac}
\newcommand{\s}{\sum}

\newcommand{\rro}{\right)}
\newcommand{\lro}{\left( }
\newcommand{\lsq}{\left[}
\newcommand{\rsq}{\right]}

\newcommand{\be}{\begin{equation}}
\newcommand{\ee}{\end{equation}}
\newcommand{\bi}{\begin{itemize}}
\newcommand{\ei}{\end{itemize}}
\newcommand{\ben}{\begin{enumerate}}
\newcommand{\een}{\end{enumerate}}

\newcommand{\rv}{{\bf r}}

\newcommand{\ej}{\epsilon_j}
\newcommand{\beq}{\begin{equation}}
\newcommand{\eeq}{\end{equation}}
\newcommand{\bea}{\begin{eqnarray}}
\newcommand{\eea}{\end{eqnarray}}

\newcommand{\<}{\langle}
\renewcommand{\>}{\rangle}
\renewcommand{\(}{\left(}
\renewcommand{\)}{\right)}
\renewcommand{\[}{\left[}
\renewcommand{\]}{\right]}
\newcommand{\commentout}[1]{{}}
\newcommand{\half}{\hbox{$1\over2$}}
\newcommand{\eq}[1]{Eq.~\eqref{#1}}

\begin{document}
\title{Quantum dynamics in splitting a harmonically trapped Bose-Einstein condensate\\ by
an optical lattice: Truncated Wigner approximation}
\author{L. Isella}
\affiliation{European Commission, Joint Research Centre, I-21020
Ispra (Va), Italy}
\author{J. Ruostekoski}
\affiliation{School of Mathematics, University of Southampton,
Southampton, SO17 1BJ, UK}

\begin{abstract}
We study the splitting of a harmonically trapped atomic
Bose-Einstein condensate when we continuously turn up an optical
lattice (or a double-well) potential. As the lattice height is
increased, quantum fluctuations of atoms are enhanced. The resulting
nonequilibrium dynamics of the fragmentation process of the
condensate, the loss of the phase coherence of atoms along the
lattice, and the reduced atom number fluctuations in individual
lattice sites are stochastically studied within the truncated Wigner
approximation. We perform a detailed study of the effects of
temperature and lattice height on atom dynamics, and investigate the
validity of the classical Gross-Pitaevskii equation in optical
lattices. We find the atom number squeezing to saturate in deep
lattices due to nonadiabaticity in turning up of the lattice
potential that is challenging to avoid in experiments when the
occupation number of the lattice sites is large, making it difficult
to produce strongly number squeezed (or the Mott insulator) states
with large filling factors. We also investigate some general
numerical properties of the truncated Wigner approximation.
\end{abstract}

\date{\today}
\maketitle

\section{Introduction}

The experimental progress in loading ultra-cold atomic gases in
weakly coupled mesoscopic traps formed by periodic optical
potentials has provided a dilute atomic system with strongly
enhanced interactions. Examples of this progress include experiments
on the Bose-Einstein condensate (BEC) coherence
\cite{AND98,GRE01,GRE02,HAD04,TUC05}, the atom number-squeezed
states \cite{ORZ01}, the Mott insulator (MI) phase transition
\cite{gremott,mott1d,tonks,XU05,FER05}, atom dynamics
\cite{BUR01,MOR01,FAL04,FER05,SAR05}, and fermionic systems
\cite{MOD03,KOH05,CHI06}. Ultra-cold atoms trapped in an optical
lattice resemble traditional condensed matter crystal lattice
systems \cite{JAK05}, but optical lattices are amenable to much
higher experimental control. The lattice strength can be easily
modified \cite{GRE01,ORZ01,gremott,mott1d,tonks,GER06} and it could
be possible to engineer, e.g., complex lattice geometries
\cite{SANT04}, interatomic interactions \cite{SANP04,BUC05}, and the
spatial profiles of the atomic hopping amplitude between the lattice
sites \cite{RUO02,ROT03,JAK03,PAC04,JAV03,OST05}.

In order to produce atomic lattice systems close to their ground
state, atoms need to be adiabatically loaded into the optical
lattices by continuously turning up the lattice potential. The BEC
is initially confined in a harmonic trap. The increasing strength of
the lattice potential reduces the tunneling amplitude between the
neighboring lattice sites and the system becomes more strongly
interacting. The strong interactions enhance quantum fluctuations,
eventually destroying the long-range phase coherence of the atoms
and fragmenting the BEC. If the turning-up of the lattice potential
is not adiabatic, the quantum fluctuations of the atoms can be far
from their ground state properties, resulting in complex dynamics.

In order to preserve adiabaticity during the turning-up of the
optical lattice potential, the rate of change of the Hamiltonian has
to be slower than any time scale of the system. The required slow
ramping-up time generally makes it very difficult to reach the MI
ground state experimentally, when there are many atoms per lattice
site, as we demonstrated in Ref.~\cite{ISE05}. The situation is
different from the MI state experiments
\cite{gremott,mott1d,tonks,XU05,FER05} with small filling factors,
because of rapidly emerging energy gap and larger quantum
fluctuations in the case of small atom numbers. The nonadiabaticity
of the turning up of the optical lattice has recently been
theoretically studied also by using classical Gross-Pitaevskii
equation (GPE) (without thermal and quantum fluctuations): The
buildup of an inhomogeneous phase profile \cite{BAN02,MCK05} was
identified in the classical analysis as a potential signature for
nonadiabaticity.

In this paper we study matter wave dynamics beyond the classical
mean-field theory by considering a harmonically trapped BEC that is
dynamically split by an optical lattice potential. In our numerical
model the quantum and thermal fluctuations of the atoms are included
in a classical stochastic field description within the truncated
Wigner approximation (TWA). We provide a more detailed and extended
analysis of our previous results \cite{ISE05} by also studying the
limits of validity of the classical GPE and the general numerical
properties of the TWA. The lattice systems we consider have large
occupation numbers per site. Recent experimental studies by Orzel
{\it et al.}~\cite{ORZ01} on the squeezing of atom number
fluctuations have been performed in the similar limit of strong
optical lattices and large filling factors. The states with squeezed
atom number fluctuations with high occupation numbers are of great
interest in precision measurements, as they may allow Heisenberg
limited interferometry with potential technological applications
\cite{HOL93}.

We study the nonequilibrium dynamics of the BEC that is split by the
lattice potential, describing the loss of phase coherence along the
lattice, the atom number fluctuations within individual lattice
sites, and the fragmentation process of the BEC. We show in
numerical TWA simulations how the atom number fluctuations evolve in
time after the turning up of the lattice potential from almost
Poisson initial value to a strongly reduced result that depends on
the parameters of the system. We also show that the squeezing of
atom number fluctuations saturates for deep lattices, as a
consequence of nonadiabatic turning up of the lattice potential. The
saturation of the number squeezing for deep lattices was
experimentally observed in Ref.~\cite{ORZ01}. This system was not
tightly elongated as our 1D numerical model, but we can still make
qualitative comparisons to the experimental data. Although the
saturation was assumed in Ref.~\cite{ORZ01} to be an artifact of the
analysis method of the interference measurement, we also numerically
find a similar saturation effect in a qualitative agreement with the
experiment. Since the atom number fluctuations in Ref.~\cite{ORZ01}
were indirectly detected from the phase noise of the interference
measurement, not all the experimentally detected phase fluctuations
may have directly corresponded to the atom number squeezing, due to
the possibility of nonadiabatic loading of the atoms. We find,
however, in the TWA simulations that considerable atom number
squeezing can be present even in the nonadiabatic regime in the
presence of increased phase noise and a nonuniform phase profile.

In the numerical TWA simulations we first consider a BEC that is
split by a simple double-well potential and compare the numerical
results to the phase collapse rate calculated for an equilibrium BEC
in a double-well potential. We then apply the TWA to the splitting
of a BEC by a periodic optical lattice potential that constitutes
the main results of the paper. We evaluate the dynamics of the phase
coherence between atoms occupying different lattice sites and the
atom number fluctuations in individual sites during and after the
the splitting of the BEC. We study in detail different cases when we
vary the final height of the lattice potential, the initial
temperature and nonlinearity. In order to investigate the limits of
validity of the classical GPE, we also systematically compare the
TWA simulation results to the GPE results by varying the initial
temperature and the final lattice height. In shallow lattices and at
low temperatures the two approaches yield very similar results for
the coherence properties and atom statistics. However, as the
lattice height and/or the temperature are increased the validity of
the GPE becomes poorer.

We also study the generation of the initial state in the TWA
simulations using an ideal gas. The stochastic noise is sampled for
noninteracting atoms, before turning up the nonlinearity in order to
produce the initial state of an interacting system. We find that
this technique works better at low temperatures in which case the
results are close to the ones obtained by more accurate TWA
calculations.

Since the TWA simulations return symmetrically order expectations
values, we also investigate the importance of the particular
stochastic representation of the density matrix. We demonstrate by
numerical examples that the correct transformation between symmetric
and normal operator orderings is very crucial in obtaining the
correct physical results, especially at low temperatures.

The TWA was introduced in nonlinear optics to study quantum
fluctuations \cite{DRU93} and it has provided successful
descriptions for nonlinear optical squeezing \cite{COR01}. The TWA
and other closely related classical field methods
\cite{ISE05,Steel,SIN01,SIN02,GAR02,GOR02,POL03,POL03c,RUO05,DAV02,DUI02,DAB06}
have previously also been applied to the studies of atomic BEC
dynamics. In our earlier studies \cite{ISE05,RUO05} we found that
the TWA is particularly useful to study bosonic atom dynamics in
optical lattices in the limit where the full multi-mode dynamics,
beyond the tight-binding approximation, becomes important and when
the atom filling factor in the lattice is large. In
Ref.~\cite{RUO05} the TWA was able to describe the experimentally
observed dissipative atom dynamics in tightly confined, shallow 1D
optical lattices \cite{FER05} even in a strongly fluctuating quantum
system.

In Sec.~\ref{overview} we introduce the TWA and the methods to
extract the normally ordered physical observables in the numerical
simulations. We also describe the general numerical approach used in
the TWA simulations in Sec.~\ref{numapproach}. The splitting of a
harmonically trapped BEC by an optical potential is studied in
Sec.~\ref{split}. We first consider a double-well potential in
Sec.~\ref{dble} and an optical lattice with a small number of sites
in Sec.~\ref{toy-system}. The main numerical results of the paper
are presented in Sec.~\ref{numerics} where we address an optical
lattice with a larger number of sites. The effects of the final
lattice height on the atom number fluctuations and matter wave
coherence are studied in Sec.~\ref{adiabaticity}. We analyze the
adiabaticity of the turning up of the lattice potential and present
qualitative comparisons to the experiment of Ref.~\cite{ORZ01}. The
effects of initial temperature and nonlinearity on the dynamics are
studied in Sec.~\ref{initemp}. We compare the TWA simulation results
to the classical GPE results in Sec.~\ref{valGPE} by varying the
initial temperature and the final lattice height. The change of
temperature during the splitting is analyzed in Sec.~\ref{ret-lat}.
An alternative method for generating the initial state of the TWA
simulations by turning up the nonlineariaty is studied in
Sec.~\ref{ramp-non-lin}. In Sec.~\ref{grnoise} we consider the
effect of the initial state noise sampling in the TWA and in
Sec.~\ref{ordering} the importance of symmetric operator ordering in
the Wigner representation. A few concluding remarks are made in
Sec.~\ref{remarks}. The Bogoliubov approximation in a discrete
tight-binding approximation is introduced in
Appendix~\ref{bogo-approach} and the three-body losses are estimated
in Appendix~\ref{3body}.

\section{Overview of the Method} \label{overview}
\subsection{Truncated Wigner approximation}

We assume that the bosonic atoms are trapped in a tight elongated
(along the $x$ direction) cigar-shaped (prolate) trap
$V_{3D}(x,y,z)=m[\omega^2x^2+\omega_\perp^2(y^2+z^2)]/2$, with the
trap frequencies satisfying $\omega\equiv\omega_x\ll
\omega_\perp\equiv\omega_y=\omega_z$. Here the radial frequency is
denoted by $\omega_\perp$ and the axial frequency by $\omega$. We
ignore the density fluctuations along the transverse directions in
order to obtain an effective 1D Hamiltonian:
\begin{align}\label{hamiltonian}
\h H=&\i dx\, \h\psi^\d(x) \h{\mathcal{L}}\h\psi(x) \nonumber
\\ &+ \f{g_{1D}}{2}\i dx\,
\h\psi^\d(x)\h\psi^\d(x)\h\psi(x)\h\psi(x)\,.
\end{align}
Here
\beq
\h{\mathcal{L}}\equiv \h{T}+V_h(x)+V_o(x,t)-\mu\,,
\label{singleparticle}
\eeq
includes the kinetic energy $\h T\equiv-\hbar^2\partial_x^2/(2m)$,
the harmonic trapping potential along the axial direction $V_h\equiv
m\omega^2x^2/2$, and the time-dependent periodic optical lattice
potential, or in the case of the double-well a Gaussian potential
barrier, $V_o(x,t)$. The atom mass and the chemical potential are
denoted by $m$ and $\mu$, respectively. The effective 1D interaction
strength is given by $g_{1D}=2\hbar\omega_\perp a$, where $a$ is the
scattering length.

We study the dynamics of a finite-temperature BEC within the TWA
when we load the atoms into an optical lattice by ramping up the
periodic lattice potential. The TWA may be obtained by using the
familiar techniques of quantum optics \cite{walls,GAR} to derive a
generalized Fokker-Planck equation for the Wigner distribution of
the trapped multi-mode BEC~\cite{Steel}. We obtain from the
Hamiltonian (\ref{hamiltonian}):
\begin{align}\label{FP}
\f{\p W(\psi,\psi^*)}{\p t}=\i_{-\infty}^{\infty}& dx\, i\Bigg{\{}
\f{\delta}{\delta\psi}\lsq \(
\h{\mathcal{L}}+g_{1D}(\v\psi\v^2-1)\)\psi\rsq
\nonumber \\
& \mbox{}-
\f{1}{4}\f{\delta^3}{\delta^2\psi\delta\psi^*}\psi\Bigg{\}}
W(\psi,\psi^*)+ \rm{c.c.}
\end{align}
Here the density operator of the quantum system is represented by a
classical quasidistribution function $W(\psi,\psi^*)$ of the complex
functions $(\psi,\psi^*)$ that correspond to the field operators
$(\h\psi,\h\psi^\dagger)$. The expectation values in the Wigner
representation $\<\cdots\>_W$ are obtained from the
quasidistribution function:
\begin{align}
&\<\psi^*(x_1)\cdots\psi^*(x_k)\psi(x_{k+1})\cdots\psi(x_l) \>_W=\nonumber \\
&\int d^2\psi\, W(\psi,\psi^*)
\psi^*(x_1)\cdots\psi^*(x_k)\psi(x_{k+1})\cdots\psi(x_l)\,.\label{wignerexp}
\end{align}
The expectation values obtained according to the Wigner distribution
correspond to the expectation values of quantum operators that are
in symmetric, or Weyl, order.

The diffusion matrix of the Fokker-Planck equation \eqref{FP} for
$W(\psi,\psi^*)$ vanishes identically and the dynamical quantum
noise acts via third-order derivatives. It is useful to write
\eq{FP} in terms of stochastic differential equations for
$(\psi,\psi^*)$ whose ensemble average of the dynamics generates the
expectation values \eqref{wignerexp} obtained from the
quasidistribution function. The TWA consists of neglecting the
third-order derivatives in \eq{FP}, resulting in a deterministic
equation for the classical field $\psi_W$ which coincides with the
Gross-Pitaevskii equation (GPE)~\cite{Steel}:
\begin{equation} \label{GPE}
i\hbar{\partial\psi_W( x,t)\over \partial t}=\[
\hat{\mathcal{L}}+g_{1D}(\v\psi_W(x,t)\v^2-1)\] \psi_W(x,t)\,.
\end{equation}
Here we have introduced a subscript in $\psi_W$ in order to
emphasize that it denotes the classical Wigner representation of the
field operator. We have also explicitly included the constant phase
factors from \eq{FP}. Although the time evolution represented by
Eq.~(\ref{GPE}) is deterministic, the thermal and quantum
fluctuations are still included in the initial state of $\psi_W$
that represents an ensemble of Wigner distributed wave functions.
The neglected terms in the TWA are small when the amplitudes of the
Wigner distribution are large. It should be emphasized that without
ignoring the third-order derivatives in the Fokker-Planck equation
no simple stochastic description exists. The stochastic
representation in the TWA for the field operator exists for the
states for which $W(\psi_W,\psi_W^*)$ is positive. The stochastic
description is especially useful since a single field $\psi_W$
incorporates both the condensate and the noncondensate populations.

The atoms are initially assumed to be in thermal equilibrium in a
harmonic trap, before the periodic optical lattice potential is
turned up. In order to sample the initial state stochastically, we
solve the quasiparticle excitations of the BEC within the Bogoliubov
approximation. We expand the field operator $\h\psi(x,t=0)$ in terms
of the BEC ground state amplitude $\h\alpha_0\psi_0$, with
$\<\h\alpha_0^\dagger\h\alpha_0\>=N_0$, and the excited states:
\beq
\label{field} \h\psi(x)=\psi_0(x)\h\alpha_0+ \sum_{j>0} \big[
u_j(x)\h\alpha_j-v^*_j(x)\h\alpha^\d_j \big]\,.
\eeq
Here $\h\alpha_j$ are the corresponding quasiparticle annihilation
operators, with
\beq
\<\h\alpha_j^\dagger \h\alpha_j\>=\bar{n}_j\equiv {1\over\exp{(
\ej/k_BT)}-1}\,,
\eeq
and $\psi_0$ is ground state solution of the GPE with the chemical
potential $\mu$. The quasiparticle mode functions $u_j(x)$ and
$v_j(x)$ ($j>0$) and the corresponding eigenenergies $\ej$ are
obtained as solutions to the Bogoliubov equations in the subspace
that is orthogonal to the ground state wave function $\psi_0$:
\begin{align}
\label{Bogo} \lro \h{\mathcal{L}}+2N_0g_{1D}|\psi_0|^2\rro
u_j-N_0g_{1D}\psi_0^2 v_j
& = \ej u_j,\nonumber\\
\lro \h{\mathcal{L}}+2N_0g_{1D}|\psi_0|^2\rro
v_j-N_0g_{1D}\psi_0^{*2} u_j & =-\ej v_j\,.
\end{align}

In the TWA we unravel the dynamics into individual stochastic
trajectories where the classical representation $\psi_W(x)$ of the
initial state of the quantum field operator $\h\psi(x,t=0)$ in
\eq{field} is sampled according to its Wigner distribution
$W(\psi_W,\psi_W^*)$, so that the ensemble average of these
individual realizations synthesizes the correct quantum statistical
correlation functions according to \eq{wignerexp}. In particular, in
order to construct $\psi_W(x,t=0)$ from $\h\psi(x,t=0)$ in
\eq{field}, we replace the excited state quantum operators
$(\h\alpha_j,\h\alpha_j^\d)$ (for $j>0$) by the complex random
variables $(\alpha_j,\alpha_j^*)$, obtained by sampling the
corresponding Wigner distribution of the quasiparticles. Within the
Bogoliubov approximation the operators $(\h\alpha_j,\h\alpha_j^\d)$
behave as a collection of ideal harmonic oscillators whose Wigner
distribution in a thermal bath may be easily evaluated~\cite{GAR}:
\beq
\label{wigner} W(\alpha_j,\alpha_j^*)=\f{2}{\pi}\tanh \( \xi_j\)
\exp\[ -2|\alpha_j|^2\tanh\( \xi_j\)\]\,,
\eeq
where $\xi_j\equiv \ej /2k_B T$. The Wigner function is Gaussian
distributed with the width $\bar{n}_j+\half$. The nonvanishing
contribution to the width at $T=0$ for each mode represents the
quantum noise. Since the Wigner function returns symmetrically
ordered expectation values, we have
\beq
\<\alpha_j^*\alpha_j\>_W =\int d^2\alpha_j\, W(\alpha_j,\alpha_j^*)
|\alpha_j|^2= \bar{n}_j+{1\over2} \,,
\eeq
and similarly $\<\alpha_j\>_W=\<\alpha_j^*\>_W=\<\alpha_j^2\>_W=0$,
etc. In many realistic physical situations the number of modes
required to generate the fluctuations in the initial state can
significantly exceed the lowest energy band in optical lattices
\cite{RUO05}, emphasizing the multi-band nature of the TWA.

We consider large BECs with $N_0\gg1$. Since the BEC is initially
weakly-interacting and confined in a harmonic trap with no optical
lattice, the main contribution to the matter wave coherence is due
to the thermal and quantum fluctuations of low-energy phonons that
are more important than the quantum fluctuations of the initial
state of the BEC mode. In most cases studied in the paper, we sample
the quantum noise of the BEC mode according to the Wigner
distribution of the coherent state~\cite{GAR}:
\beq \label{wc}
W_c(\alpha_0,\alpha_0^*)={2\over\pi} \,\exp\lro -2\v\alpha_0-
N_0^{1/2}\v^2\rro \,,
\eeq
for which $\langle \alpha_0\rangle_W=N_0^{1/2}$ and
$\<\alpha_0^*\alpha_0\>_W =N_0+\half$. Since we compare the matter
wave coherence between the atoms in different lattice sites, the
global BEC phase is unimportant. The advantage of using the coherent
state description is that the corresponding Wigner distribution is
positive. As shown later in the paper, the assumption of the
coherent state for the BEC mode is not very important. In fact, we
could even treat the BEC mode classically without significantly
affecting the results. In Section \ref{grnoise} we compare the
sampling of the BEC mode according to Eq.~(\ref{wc}) to the case in
which the BEC mode is treated classically having a fixed number of
atoms. A classical treatment of the ground state does not affect the
prediction for the phase coherence along the lattice, but produces
slightly smaller atom number fluctuations in the lattice sites.

Our TWA model for the trapped atomic vapor is for the Hamiltonian
describing a closed system. In the lattice experiments the atoms are
also coupled to environment, resulting in dissipation with the
system relaxing towards its ground state. We could introduce a more
sophisticated model, e.g., by incorporating the three-body losses
and the spontaneous emission due to the lattice lasers. This would
introduce also a dynamical noise term in Eq.~(\ref{GPE}). Although
the spontaneous emission generates an important loss mechanism for
the atoms in several experimental situations, it can be reduced by
further detuning the lattice laser light from the resonance of the
atomic transitions. For instance, with intense CO$_2$ lasers, that
are far off-resonant, the spontaneous emission rate can be very low
\cite{OHA99}. In Appendix~\ref{3body} we estimate the importance of
the three-body losses of atoms in a 1D optical lattice for a typical
set of parameters considered in the TWA simulations.

\subsection{Symmetric ordering} \label{interest}

The Wigner distribution returns symmetrically ordered expectation
values for the field operators. This means that in the TWA
simulations the expectation values involving the full multi-mode
Wigner fields are symmetrically ordered with respect to {\it every}
mode. In general, this can significantly complicate the analysis of
the numerical results from the TWA simulations. For instance, for
the initial state (\ref{field}) the simple spatial correlation
function in the Wigner representation would actually return:
\begin{align}
\label{correlation} \<\psi^*_W (x)\psi_W&(x')\>_W =
\<\h\psi^\dagger(x)\h\psi(x')\>+
\f{N_0}{2}\psi_0^*(x)\psi_0(x')\nonumber\\
&+{1\over2} \s_{i=1}[u_i^*(x)u_i(x')+v_i(x)v_i^*(x')]\,,
\end{align}
where $\<\cdots\>_W$ denotes the expectation value obtained from the
TWA simulations and $\<\cdots\>$ the normally ordered expectation
value of the quantum operators. In order to extract normally ordered
expectation values for the correlation functions of the BEC from the
TWA simulations, Steel {\it et al.} in Ref.~\cite{Steel} defined a
`condensate mode' operator associated with the projection of the
stochastic field onto the ground state solution. This was used to
calculate the phase diffusion of a spatially static, harmonically
trapped BEC. Since here we study the splitting of a BEC by a
periodic optical lattice potential, it is useful to define
analogously the ground state operators $a_j$ for each individual
lattice site $j$:
\beq
\label{projection} a_j(t)=\int_{j^{\rm th} {\rm well}}dx\,
\psi^*_{0}(x,t)\psi_W(x,t)\,,
\eeq
where $\psi_W(x,t)$ is the stochastic field, determined by
Eq.~(\ref{GPE}), and $\psi_0(x,t)$ is the ground state wave function
at time $t$, obtained by integrating the GPE in imaginary time in
the potential $V(x,t)$. The integration is over one lattice site. In
the following we also use the same description in a double-well
potential in which case we integrate over the left and the right
wells. The importance the correct operator ordering is demonstrated
by a numerical example in Section \ref{ordering}.

With the projection method we can avoid the complications arising
from the symmetrically ordered multi-mode field $\psi_W$. Using the
definition in Eq.~\eqref{projection}, the normally ordered
expectation values can be easily obtained for each lattice site
ground state mode $a_j$. For instance, the atom number in the ground
state in the $j$th site reads:
\beq
\label{nj} n_j=\langle \h a^\d_j\h a_j\rangle= \<a_j^*a_j\>_W
-\f{1}{2}\,,
\eeq
and the atom number fluctuations in the $j$th site:
\begin{align}\label{deltaj}
\Delta n_j=&\lsq\langle (\h a_j^\d\h a_j)^2\rangle-\langle\h
a_j^\d\h a_j\rangle^2\rsq^{1/2} \nonumber \\ =& \lsq\langle ( a_j^*
a_j)^2\rangle_W-\langle a_j^* a_j\rangle_W^2-\f{1}{4}\rsq^{1/2}\,.
\end{align}
Similarly, in order to characterize the phase coherence along the
lattice, we may introduce the normalized first-order correlation
function $C_j$ between the atoms in the central well of the lattice
and in its $j$th neighbor:
\beq \label{coherence-def}
C_j=\f{|\langle \h a_0^\d \h a_j\rangle|}{\sqrt{n_0n_j}}\,.
\eeq

In order to obtain directly the fluctuations in the relative phase
operator $\h\varphi_{0j}\equiv \h\varphi_j-\h\varphi_0$ between the
atoms in the central lattice site and in its $j$th nearest neighbor,
we evaluate  $D_j$ defined as:
\beq
D_j\equiv|\<e^{[i(\h\varphi_0-\h\varphi_j)]}\>|\,.\label{bigD}
\eeq
Here $\h a_j=|\h a_j|e^{\h\varphi_j}$ and $D_j$ is calculated by
normalizing $\h a_0^\d \h a_j$ in each stochastic trajectory before
the averaging. Typically $D_j$ and $C_j$ yield almost equal results
and we usually only show $C_j$.

\subsection{Numerical approach}\label{numapproach}

We study the nonequilibrium quantum dynamics of the bosonic atoms
within the TWA. The BEC, which is initially confined in a harmonic
trap, is split by a periodic multi-well optical lattice or a
double-well potential. The atoms are initially assumed to be in
thermal equilibrium and we first solve the ground state amplitude
profile $\psi_0(x)$ by evolving the GPE in imaginary time in the
harmonic trap. We then diagonalize the Bogoliubov equations
[Eq.~(\ref{Bogo})] in the subspace orthogonal to $\psi_0(x)$ in
order to obtain the quasiparticle eigenfunctions $u_j(x),v_j(x)$ and
the corresponding eigenenergies $\ej$.

Throughout the paper we use large atom numbers, $N_0=2000$ atoms at
the beginning of the ramp, so that the occupation number of the
central lattice sites is always high. For the typical nonlinearity
$N_0g_{1D}=100\hbar\omega l$, with $l\equiv (\hbar/m\omega)^{1/2}$,
the corresponding initial Thomas-Fermi radius
$R/l=(3N_0g_{1D}/2\hbar\omega l)^{1/3}\simeq5.3$. In 1D the strength
of interactions is commonly expressed in terms of the ratio $\gamma$
between the interaction energy and the kinetic energy needed to
localize the atoms within the mean interatomic distance $1/n_{1D}$,
where $n_{1D}$ is the 1D atom density \cite{KHE03}. We obtain
$\gamma=mg_{1D}/(\hbar^2n_{1D})\alt 10^{-3}$ and the initial
harmonically trapped BEC is well described by the GPE and the
Bogoliubov theory.

The time evolution of the ensemble of Wigner distributed
wavefunctions, determined by Eq.~(\ref{GPE}), is unraveled into
stochastic trajectories, where the initial state of each realization
for the classical stochastic field $\psi_W$ is generated using the
expansion (\ref{field}) with the operators replaced by the complex,
Gaussian-distributed [Eqs.~(\ref{wigner}) and~(\ref{wc})] random
variables $(\alpha_j,\alpha_j^*)$. During the time evolution we
continuously increase the strength of the periodic optical lattice
potential or, in the case of the double-well, a repulsive Gaussian
potential at the center of the trap. The integration of the time
dynamics [Eq.~(\ref{GPE})] is performed using the nonlinear
split-step method \cite{JAV06} on a spatial grid of up to 4096
points and in several cases with the optical lattice potential the
sufficient convergence is obtained after 600 realizations, whereas
in a double-well potential typically around 800 realizations are
required. The convergence is generally slower at higher
temperatures. Finite temperature systems require more iterations and
are therefore numerically more demanding. This is clear in the
sampling of the Wigner distribution in Eq.~\eqref{wigner} whose
width increases with the initial temperature $T_i$, indicating
higher excited level population and more noise in the initial state
Wigner distribution. Unlike the 3D TWA \cite{SIN02}, the 1D
simulations do not similarly depend on the total number of
quasiparticle modes and we found the calculated results to be
unchanged when we varied the number of modes.

\section{Splitting a BEC by an optical potential}\label{split}

\subsection{Double-well potential} \label{dble}

Before turning to the dynamical studies of bosonic atoms fragmented
by an optical lattice, it is useful first to introduce the TWA
method in a much simpler double-well potential case. Since the
projection method we found very useful in an optical lattice is much
less accurate in a double-well potential for short nonadiabatic
ramping-up times, we only present here some example cases and will
publish a more detailed study elsewhere.

The splitting of a BEC by a double-well potential has attracted
considerable theoretical interest in the past (see, e.g.,
Refs.~\cite{LEG91,JAV-dble}, and references therein). Also the
finite-temperature dynamics of atomic BECs in a double-well
potential has been investigated \cite{RUO98b,ZAP03}. Recent
experiments include, e.g., the use of a double-well BEC as a noise
thermometer \cite{GAT06}.

\begin{figure}
\includegraphics[width=0.49\columnwidth]{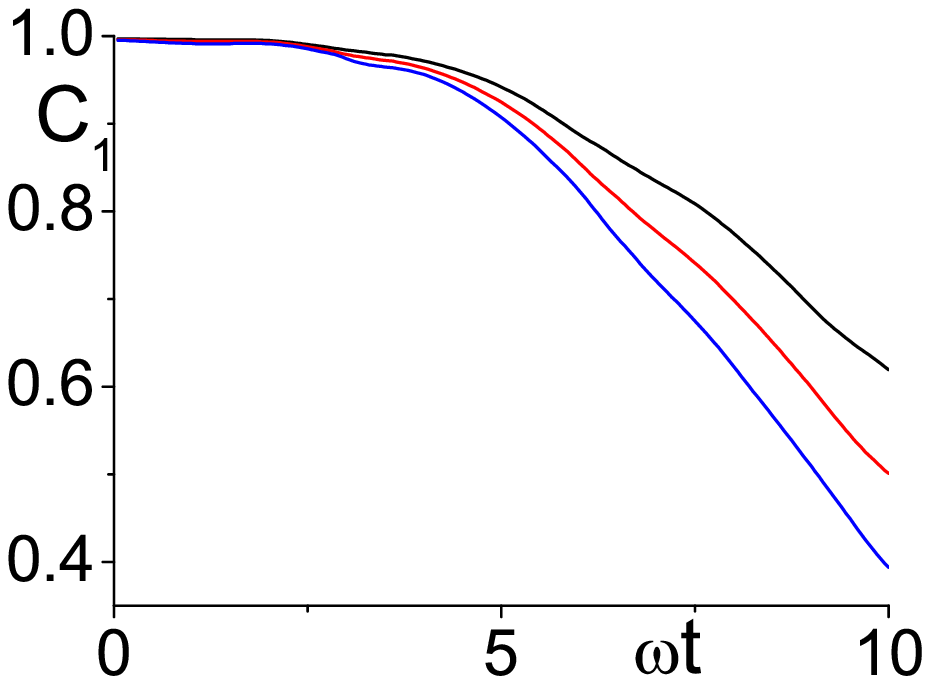}
\includegraphics[width=0.49\columnwidth]{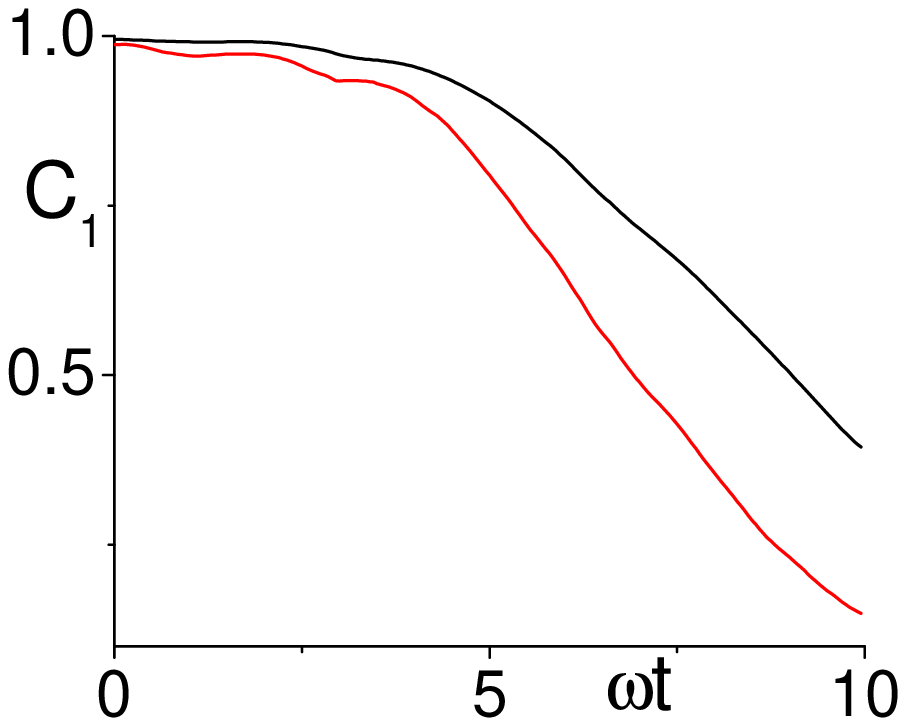}
\vspace{-0.3cm} \caption{The dynamics of the phase coherence during
the splitting of a BEC by a Gaussian potential barrier. The phase
coherence between the atoms in the two wells for different
nonlinearities (on left) $N_0g_{1D}/(\hbar\omega l)=100,150,200$
(the curves from top to bottom) at the initial temperature $T_i=0$.
The phase coherence at two different initial temperatures (on right)
$k_BT_i/\hbar\omega=0$ and $37.8$ (the lower curve) for
$N_0g_{1D}=200\hbar\omega l$. In the both cases the final value of
the Gaussian potential is $A_f=1000\hbar\omega$, and its width
$\sigma=0.5l^2$. The ramping-up time is $\omega\tau=5$ and
$N_0=2000$.}\label{figwell}
\end{figure}

We use the similar set-up as in the early BEC experiment of
Ref.~\cite{AND97}, where a harmonically trapped BEC was split by a
blue-detuned far-off resonant laser beam. We model the laser by a
Gaussian potential in \eq{singleparticle}
\beq
V_o(x,t)=A(t)\exp(-x^2/\sigma)\,,
\eeq
with $\sigma=0.5l^2$. During the time evolution, we increase
exponentially the laser potential to some final value $A_f$ at time
$\tau$ according to $A(t)=\exp(\kappa t)-1$. In Fig.~\ref{figwell}
we show the dynamics of the phase coherence $C_1$ between the two
wells, as defined in Eq.~(\ref{coherence-def}). We vary the
nonlinearity and the initial temperature. The ramping-up time of the
Gaussian potential is $\omega\tau=5$. As expected, both the increase
in the nonlinearity and in the initial temperature result in a
faster reduction of the relative phase coherence. However, the
projection to the ground state in Eq.~(\ref{projection}) in the
double-well case resulted in notable atom number oscillations and
the calculated phase coherence values should therefore be considered
only as an order of magnitude estimates. We can compare the
zero-temperature case of the collapse time of the relative phase
coherence between the atoms in the two wells to the estimates
obtained from the Bogoliubov theory for a ground state BEC
\cite{MOLMER}. If we generalize the results of Ref.~\cite{MOLMER}
for the present situation and for arbitrary $\Delta n$, we obtain
for the collapse time $\tau_c$:
\beq
\label{taumolmer} \tau_c=24^{1/3}\ln2\(\f{g_{1D}}{\hbar\omega
l}\)^{-2/3}\f{n^{1/3}}{\omega\Delta n }\,.
\eeq
After a time $\tau_c$, the value of the coherence function $C_1$ is
reduced to 0.5. The values of $n$ and $\Delta n$ in
Eq.~\eqref{taumolmer} could be obtained from the numerical results
of the TWA simulations, e.g., by averaging $\Delta n$ after the
ramping process. If we assume the atom number fluctuations between
the two wells to be Poissonian (or binomial), we obtain for the
collapse times $\omega\tau_c=3.3,2.5,2.1$, for the nonlinearities
$N_0g_{1D}/(\hbar\omega l)=100,150,200$, respectively. The collapse
time of the TWA simulations is estimated by determining the time
when $C_1=0.5$ and subtracting the ramping-up time $\omega\tau=5$
from this value. For the same nonlinearities we obtain
$\omega\tau_c>~5$, $\omega\tau_c\simeq~5$, and
$\omega\tau_c\simeq~4$.

The collapse time for the phase coherence is quite sensitive to the
atom number fluctuations and could be experimentally used to measure
$\Delta n$. The collapse time in the TWA simulations indicates
sub-Poisson atom number fluctuations. The simple collapse time
estimate from Eq.~\eqref{taumolmer}, that does not take into account
the effects of the dynamical splitting of the BEC, yields $\Delta n
\sim \Delta n_{\rm Poisson}/2$. The numerical values of $\Delta n$
in the TWA simulations are not reliable because of the simple
projection method used in the double-well problem.

\subsection{Optical lattice with a small number of sites}\label{toy-system}

The condensate fragmentation is generally much more difficult to
analyze when the number of wells is increased from two \cite{STR04}.
The TWA, incorporating quantum and thermal fluctuations for the
complete field operator, becomes a very useful tool in investigating
the dynamics of the fragmentation process. Unlike in the double-well
case, the projection method to the lowest mode in each site produces
stable and more accurate results.

In this Section we demonstrate the loss of phase coherence between
the atoms in different lattice sites, when the BEC is split by
continuously turning up an optical lattice with a small number of
sites. We evaluate the phase collapse both at $T=0$ and at finite
temperature. In the following Section we turn to a more detailed
study of the turning up process of the lattice potential when the
number of lattice sites is larger.

\begin{figure}
\begin{center}
\includegraphics[angle=0, width=0.49\columnwidth]{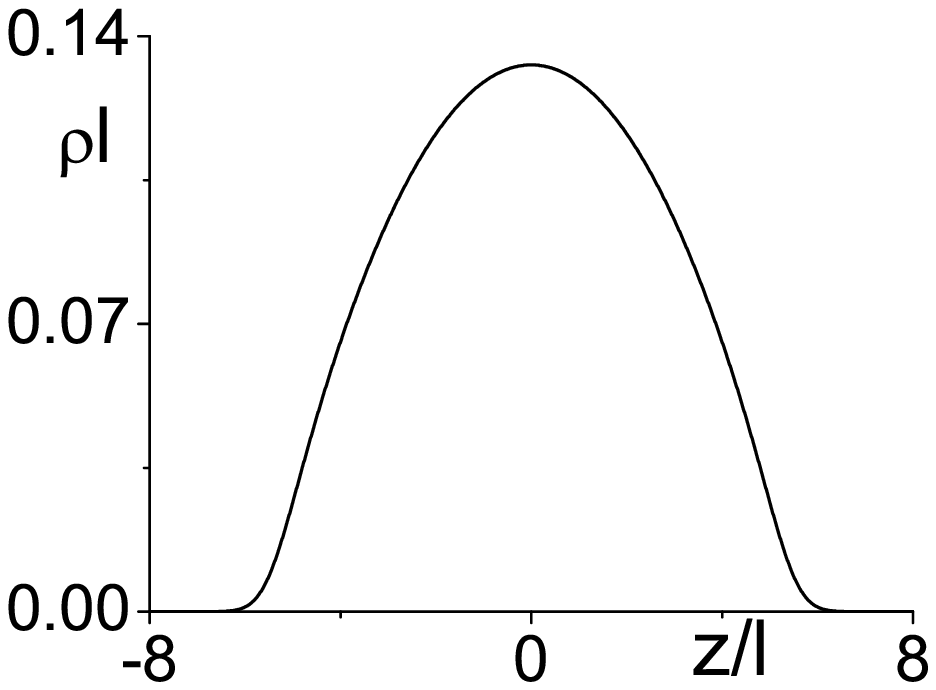}
\includegraphics[angle=0, width=0.49\columnwidth]{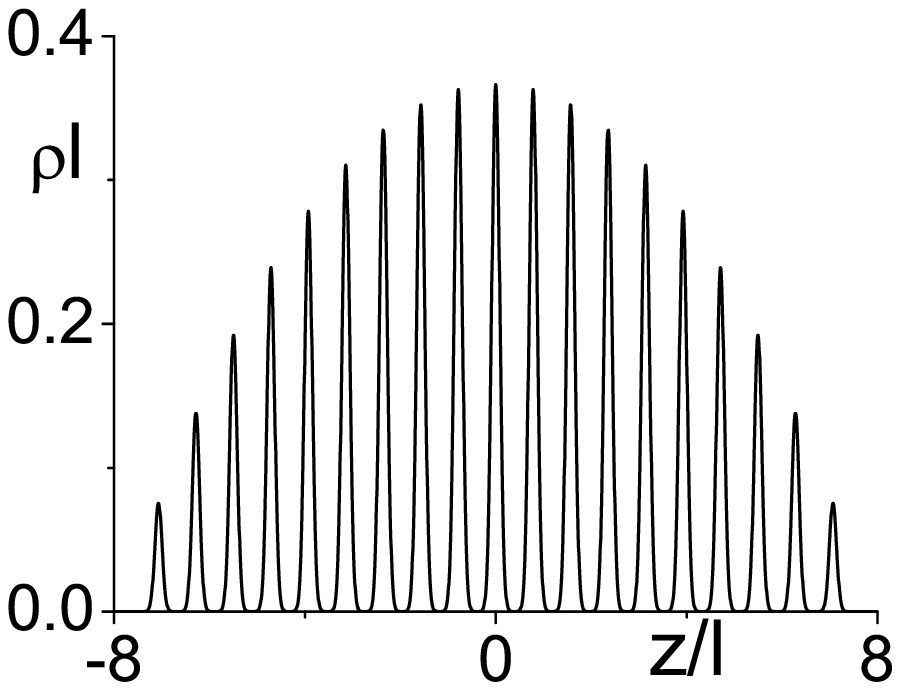}
\end{center}
\vspace{-0.4cm} \caption{The normalized ground state atom density in
a harmonic trap without optical lattice (left) and with the lattice
for $s=37.5$. The lattice has clearly pushed the atoms towards the
outer regions of the trap by increasing the radius of the atom
vapor. The nonlinearity $N_0g_{1D}=120\hbar \omega l$ and the
wavenumber $kl=4$. } \label{preliminary-dens}
\end{figure}

We start with the system in thermal equilibrium in the harmonic trap
and continuously turn up an optical lattice potential, so that
$V_o(x,t)$ in \eq{singleparticle} is defined by
\beq
V_o(x,t)=s(t) E_r \sin^2{(\pi x/d)}\,, \label{optlattice}
\eeq
where the lattice photon recoil energy
\beq
E_r\equiv{\hbar^2 \pi^2\over 2m d^2}\,,
\eeq
and $d=\lambda/2\sin(\theta/2)$ denotes the lattice period, obtained
by two laser beams intersecting at an angle $\theta$. The advantage
of 1D lattices is that the lattice spacing can be easily modified by
changing the angle between the lasers. In the simulations the height
of the lattice potential $s(t)$ is turned up exponentially during a
time $\tau$ to some final value $s$ according to
\beq
s(t)=e^{\kappa t}-1\,,
\eeq
for $t\leq\tau$. In this Section we use the lattice spacing $d=\pi
l/4$ and the nonlinearity $N_0g_{1D}=120\hbar\omega l$. For
instance, the ground state calculated at the lattice height $s=37.5$
then occupies $\sim 19$ sites; see Fig.~\ref{preliminary-dens}.
However, in a fast nonadiabatic turning up of the lattice, we study
here, only about 13-15 wells are occupied after the ramping.

As the lattice is raised the tunneling amplitude of the atoms
between the neighboring sites decreases exponentially and the system
becomes more strongly interacting. The strong interactions in the
lattice enhance quantum fluctuations, eventually destroying the
long-range phase coherence of the atoms.

The relative phase between the atoms in different lattice sites
during the turning up of the lattice potential generally indicates
the flow of atoms towards the edges of the atom cloud, as the cloud
radius increases due to the lattice. The dynamics of the phase on a
single stochastic realization also undergoes stochastic fluctuations
due to the vacuum noise. The deeper the lattice potential, the
larger are the variations in the dynamical trajectories of the phase
between individual stochastic realizations. These variations, once
averaged over a large number of realizations, result in the loss of
phase coherence.

\begin{figure}
\begin{center}
\includegraphics[angle=0, width=0.49\columnwidth]{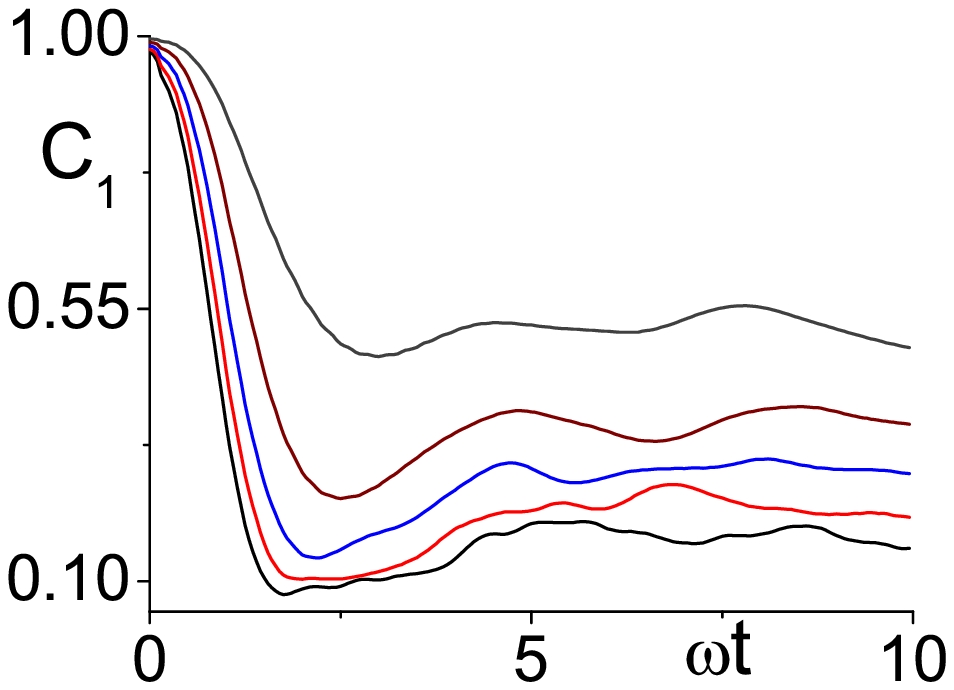}
\includegraphics[angle=0, width=0.49\columnwidth]{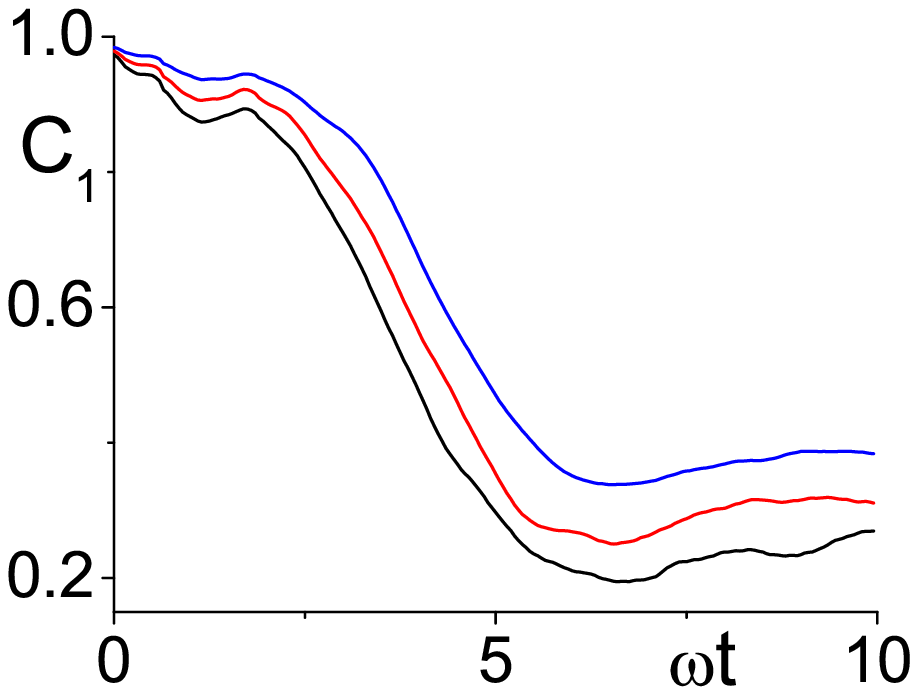}
\includegraphics[angle=0, width=0.49\columnwidth]{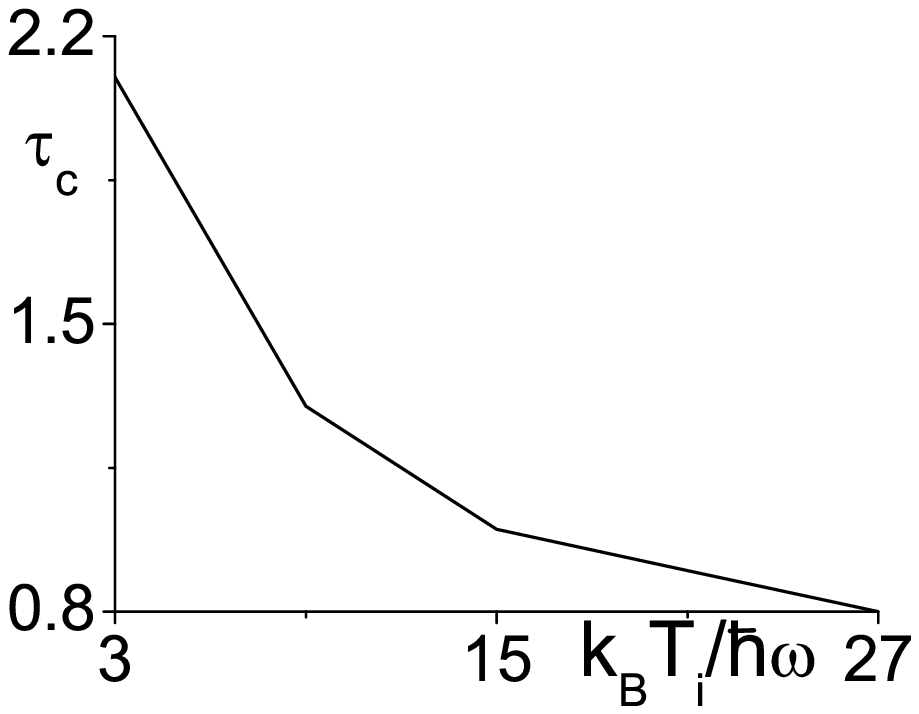}
\includegraphics[angle=0, width=0.49\columnwidth]{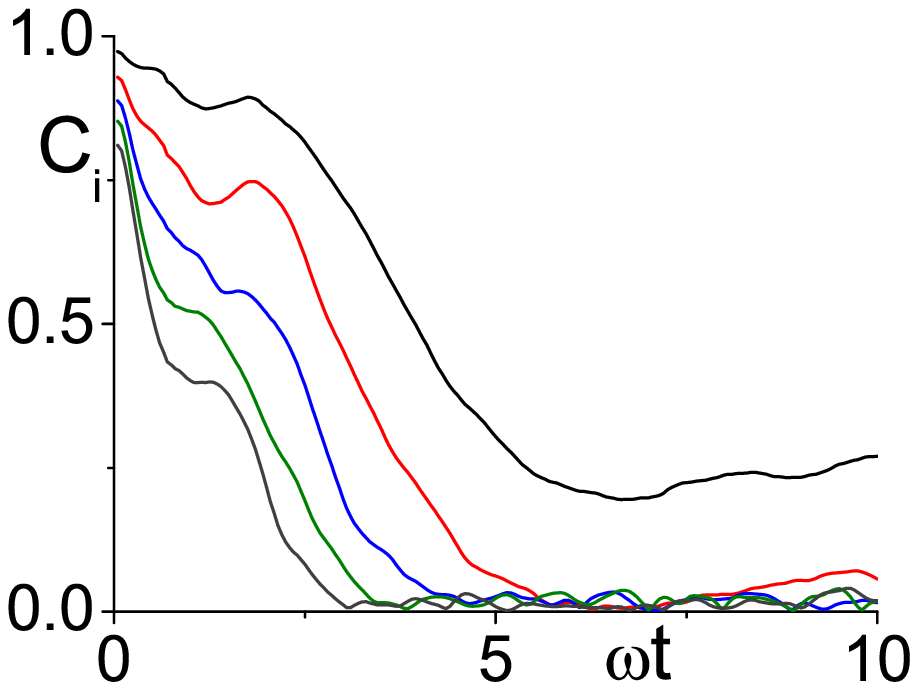}
\end{center}
\vspace{-0.4cm} \caption{The phase coherence at different
temperatures and along the optical lattice. We evaluate $C_1$ for
the ramping-up times $\omega\tau=1$ (top left; curves from top
correspond to the initial temperatures
$k_BT_i/\hbar\omega=3,9,15,21,27$) and $\omega\tau =5$ (top right;
curves from top correspond to $k_BT_i/\hbar\omega=15,21,27$). The
notch after about one trap period in the coherence function is due
to a resonance involving the first phonon modes $u_1$ and $v_1$ and
is enhanced with the increasing temperature. The estimated phase
collapse time $\tau_c$ (bottom left) as a function of the initial
temperature, obtained from the data in the top left diagram by
choosing the time when $C_1=0.65$. The coherence between the atoms
in the central site and in the first five nearest neighbors (bottom
right; curves from top $C_i$ with $i$ from one to five) for
$k_BT_i=27\hbar\omega$. In all the plots $N_0g_{1D}=120\hbar\omega
l$.} \label{firstnumerics}
\end{figure}

\begin{figure}
\begin{center}
\includegraphics[angle=0, width=0.49\columnwidth]{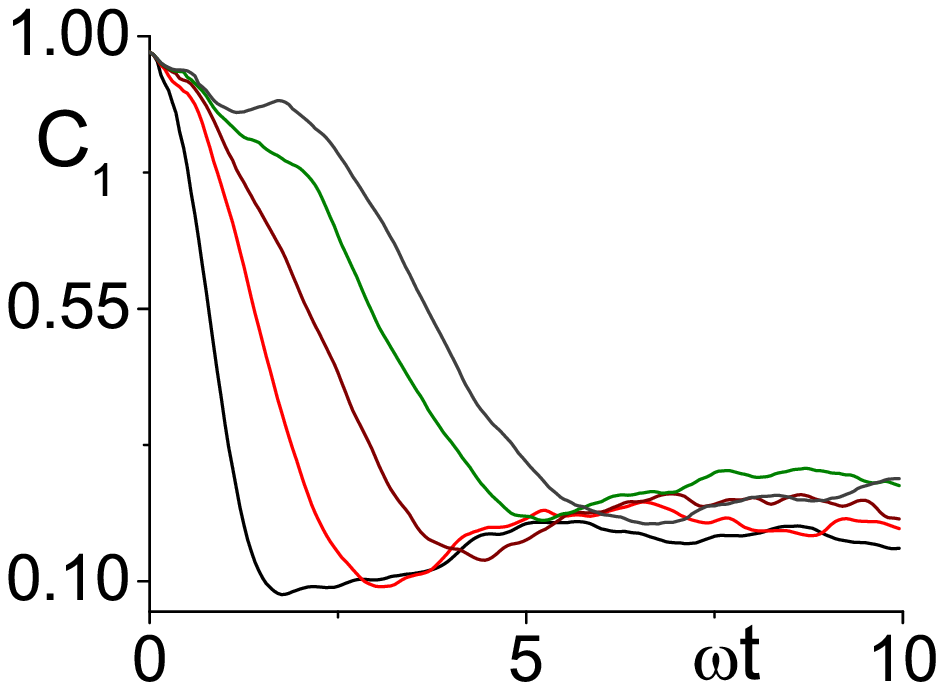}
\includegraphics[angle=0, width=0.49\columnwidth]{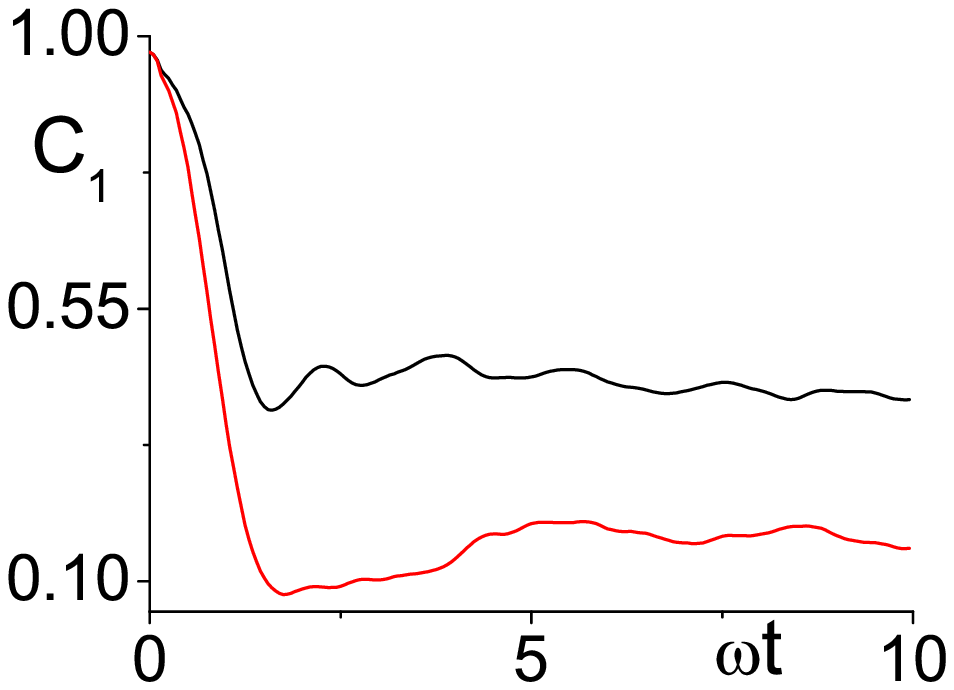}
\end{center}
\vspace{-0.4cm} \caption{The phase coherence for different
ramping-up times and lattice height. The same system as in Fig.\
\ref{firstnumerics} at $k_BT_i=27\hbar\omega$ for different
ramping-up times (on left; curves from left $\omega\tau=1,2,3,4,5$)
and for the final lattice heights $s=37.5$ (lower curve) and $s=25$
(upper curve) for $\omega\tau=1$ (on right).} \label{firstnumerics2}
\end{figure}

In Fig.~\ref{firstnumerics} we show the normalized phase coherence
$C_1$ for the atoms between the central well and its first neighbor,
as defined in \eq{coherence-def} at different initial temperatures
$T_i$, for two different values of the ramping-up time. The phase
collapse time exhibits exponential-like decrease as a function of
$T_i$ for $\omega\tau=1$. We also show the phase coherence between
the atoms in the central site and in its several nearest neighbors,
describing the phase coherence along the lattice. The effect of the
ramping-up time and the final lattice height on the phase coherence
is shown in Fig.~\ref{firstnumerics2}. It is interesting to note
that in several cases the phase coherence rapidly decays during and
immediately after the turning up of the lattice potential, but
remains surprisingly steady at some finite value afterwards.

\subsection{Optical lattice with a larger number of sites}\label{numerics}

\begin{figure}
\includegraphics[width=0.47\columnwidth]{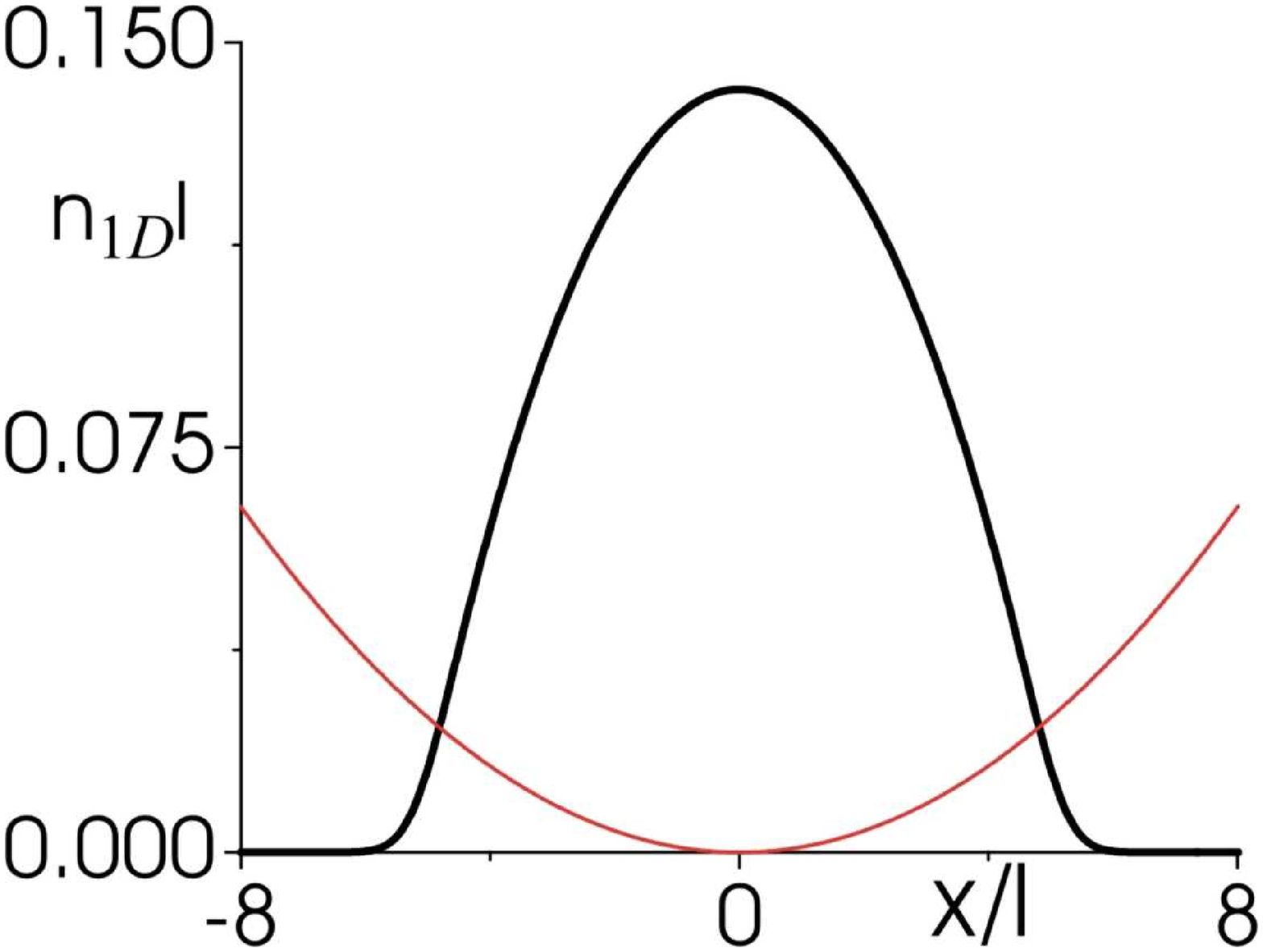}
\includegraphics[width=0.47\columnwidth]{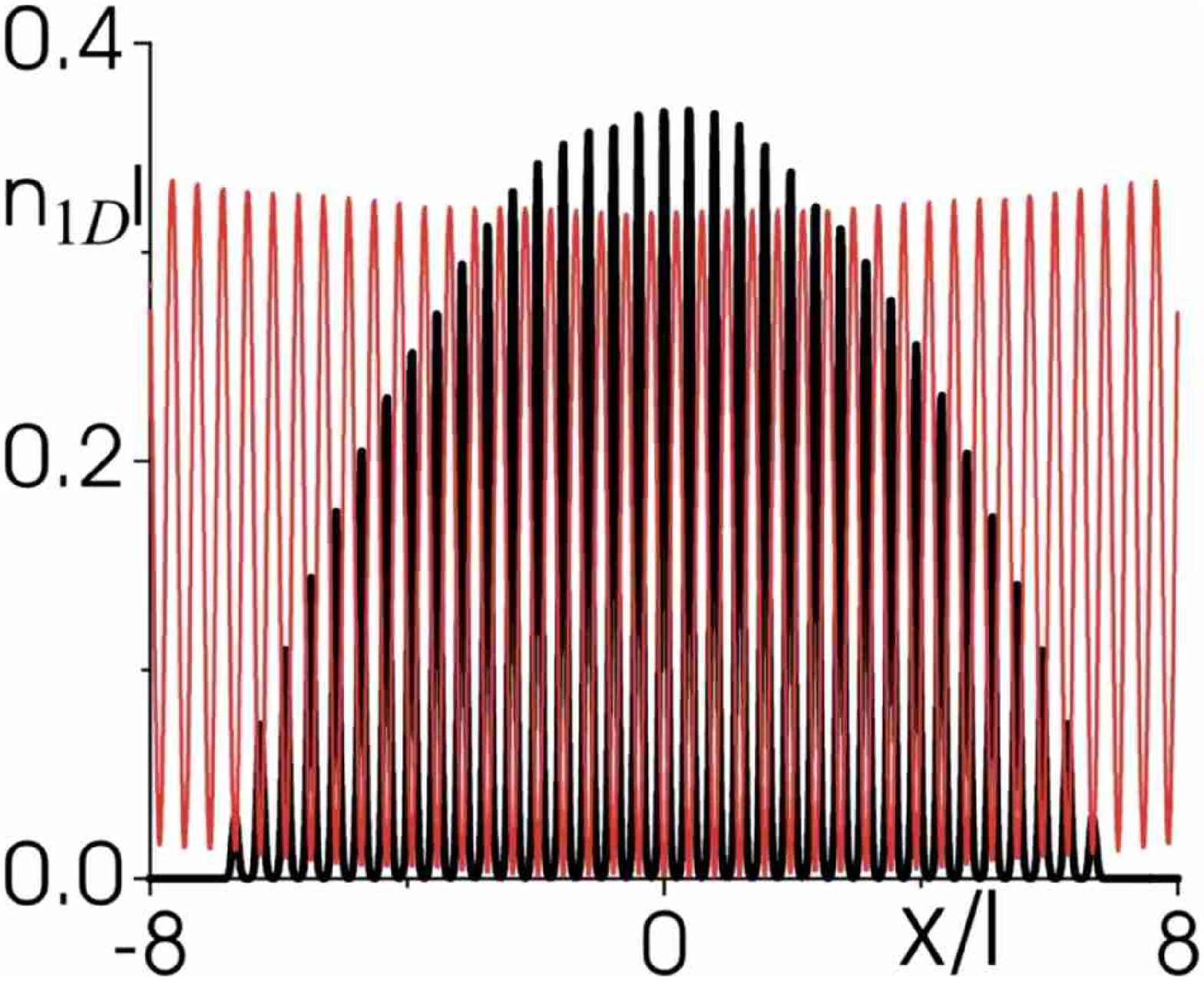}
\vspace{-0.3cm} \caption{The normalized ground state atom density
(thick line) in a harmonic (left) and in a combined harmonic and
optical lattice (right) with $s=20$. For illustrative purposes we
also show the potentials (thin line; in arbitrary units). Here
$N_0g_{1D}=100\hbar\omega l$ and $kl=8$. }\label{density}
\end{figure}

In this Section we present the main results of the paper by
considering the splitting of a harmonically trapped BEC by a rapidly
varying optical lattice potential. We choose the lattice spacing to
be $d=\pi l/8$ in \eq{optlattice}. We then have about 30-35 lattice
sites within the classical diameter $2R$ of the BEC. A similar
number of sites has also been realized in recent experiments in a
cigar-shaped trap with $d\simeq 2.7\mu$m \cite{HAD04}. The ground
state atom number $N_0=2000$ before turning up the lattice results
in the occupation number of about $n_0\simeq 90$-100 atoms in the
central site of the optical lattice. The normalized ground state
atom density is shown in Fig.~\ref{density} with (for $s=20$) and
without an optical lattice.

\subsubsection{The effect of the final lattice height}\label{adiabaticity}

In this Section we vary the final height of the optical lattice
potential when a harmonically trapped BEC is fragmented by means of
continuously turning up the lattice. We evaluate the dynamics of the
phase fluctuations between the atoms in different lattice sites and
the atom number fluctuations in individual sites. The system is
closely related to the recent experiment by Orzel {\it et al.}
\cite{ORZ01} where the atom number squeezing in strong optical
lattices with large filling factors was observed. However, the
notable difference between our model and the experimental set-up is
that in Ref.~\cite{ORZ01} the 1D optical lattice exhibited a very
weak radial confinement and did not produce a tightly-elongated 1D
gas with negligible radial excitations.

\begin{figure}
\includegraphics[width=0.49\columnwidth]{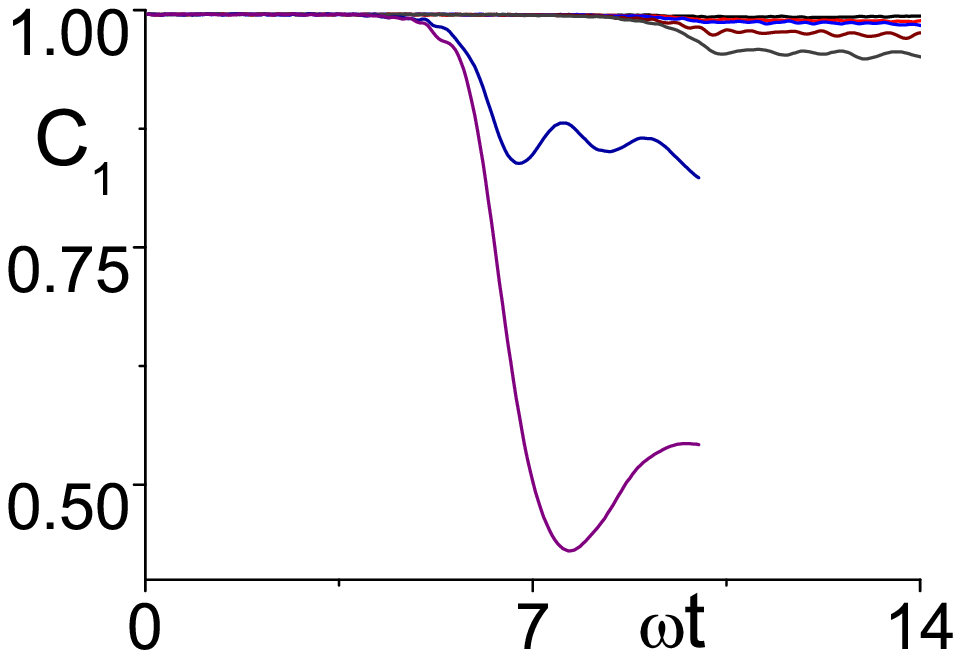}
\includegraphics[width=0.49\columnwidth]{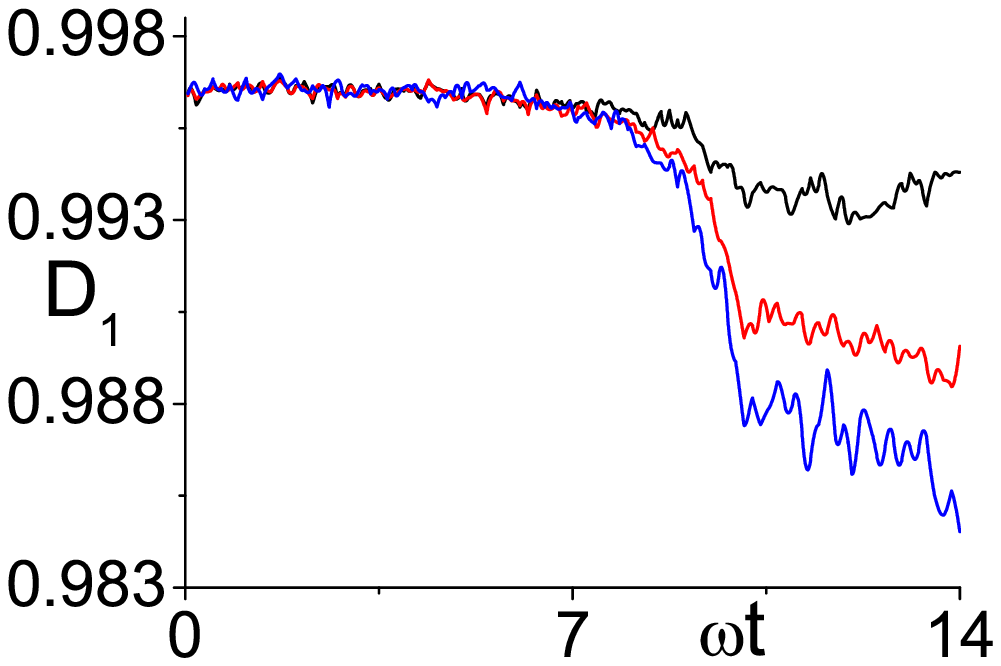}
\includegraphics[width=0.49\columnwidth]{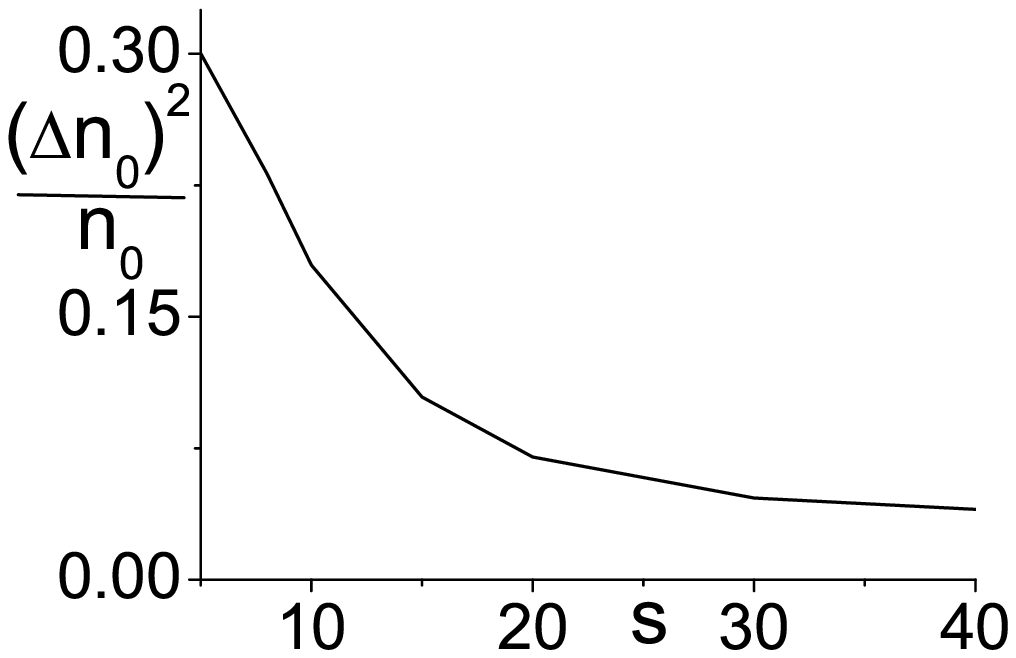}
\includegraphics[width=0.49\columnwidth]{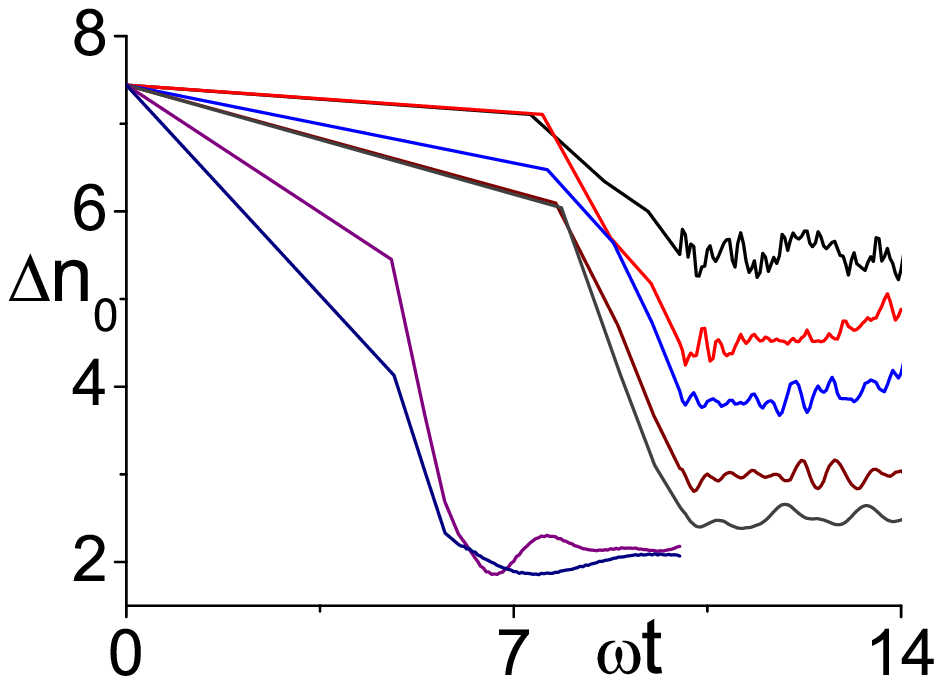}
\vspace{-0.4cm} \caption{The phase coherence between the central
well and its nearest neighbor $C_1$ as a function of time (top left)
at $T=0$ for different final heights of the optical lattice (curves
from top represent $s=5,8,10,15,20,30,40$ with the atom number in
the central well $n_0\simeq$ 90-100). The ramping-up time
$\omega\tau=10$, except for $s=30,40$, $\omega\tau=6$. Top right:
$D_1$ for $s=5,8,10$ (curves top to bottom). The number fluctuations
$\Delta n_0$ for the central well (bottom right) for the same runs.
Here $g_{1D}=0.05\hbar\omega l$ and $N_0=2000$. The number squeezing
(bottom left) as a function of the final lattice height can be
accurately fitted according to \eq{dnfit}.} \label{heightfig}
\end{figure}
In Fig.~\ref{heightfig} we show the phase coherence between the
atoms in the central well and in its nearest neighbor $C_1$
[\eq{coherence-def}] and the number fluctuations $\Delta n_0$
[\eq{deltaj}] in the central well for different final heights of the
periodic potential at $T=0$. The ramping-up time is fixed to
$\omega\tau =6$ for $s=30,40$ and to $\omega\tau =10$ for the
shallower lattices. For shallow lattices the phase coherence remains
high and steady, but for larger $s$ it is reduced and becomes
strongly oscillatory. The enhancement of phase fluctuations in
deeper lattices is associated with progressively increasing atom
number squeezing. The number squeezing can be accurately fitted
according to
\beq
{(\Delta n_0)^2\over n_0} \simeq 0.03+0.5e^{-s/8}\,.\label{dnfit}
\eeq
Due to the large occupation numbers, $\Delta n_0$ are strongly
sub-Poissonian, approaching the asymptotic value $(\Delta
n_0)^2/n_0\simeq 0.03\ll 1$ for large $s$. The numerics proved more
demanding when the final height of the optical potential reached
30-40$E_r$.

Although the discrete tight-binding Hamiltonian \eq{BH} is only
valid for weakly excited and very deep lattices, it is interesting
to compare the TWA to the Bogoliubov approximation of the BHH
\eqref{BH}, introduced in Appendix~\ref{bogo-approach}. The discrete
Bogoliubov result for the ground state in a uniform lattice
[Eq.~\eqref{deltaphi}] predicts the phase fluctuations
$\Delta\varphi_{01}$ to be weak when the hopping amplitude $J$ times
the lattice site occupation number $n$ is much larger than the
on-site interaction $U$. From the numerical results for $D_1$
[\eq{bigD}] in Fig.~\ref{heightfig} we can deduce
$(\Delta\varphi_{0,1})^2$ and for shallow lattices we find it to be
roughly by the factor of two smaller than the uniform ground state
result Eq.~\eqref{deltaphi}. The small phase fluctuations of the TWA
results may be better understood when we compare them to the phase
fluctuations obtained by solving the discrete Bogoliubov theory in a
combined harmonic trap and an optical lattice;
Fig.~\ref{num-bogolat}. Including the harmonic trap can
significantly reduce the phase fluctuations close to the trap center
and enhance them close to the edge of the atom cloud.

The Bogoliubov result of the discrete tight-binding Hamiltonian
\eq{BH} for the ground state atom number fluctuations
[Eq.~\eqref{deltan}] indicates that $\Delta n_{i}$ becomes strongly
sub-Poissonian $(\Delta n_{i})^2\ll n_i$, when $n_iU\gg J$. The
numerical TWA results for the atom number fluctuations in shallow
lattices in Fig.~\ref{heightfig} are clearly higher than the
homogeneous ground state result in Eq.~\eqref{deltan}. Moreover, in
deep lattices (see the discussion in Appendix~\ref{bogo-approach}),
the system in the ground state may undergo a quantum phase
transition from the superfluid to the Mott insulator state. Here
$n_i J\sim U$ at $s\simeq 38$. However, in the simulations we find
$\Delta n_0\agt 1$ for all $s$. The saturation of the atom number
squeezing in \eq{dnfit} and clearly higher values of $\Delta n_{0}$
than predicted for the ground state indicate that the atoms are not
loaded adiabatically into the optical lattice. The saturation can be
understood if we take into account the effects of the finite
ramping-up time of the lattice.

In order to preserve the adiabaticity during the turning-up of the
lattice potential and for the system to remain in its ground state,
we require that the rate of change of any parameter in the
Hamiltonian to be slower than any characteristic time scale of the
system. In optical lattices when the lattice height is increased,
the most rapidly changing parameter typically is the tunneling
amplitude for the atoms between neighboring sites. On the other
hand, at low lattice heights it is more difficult to avoid exciting
higher vibrational levels within one potential well, resulting in
excitations in the higher energy bands.

The phonon mode energies $\omega_j$ in the lowest energy band
decrease with increasing lattice strength [see the homogeneous
Bogoliubov result for the tight-binding Hamiltonian in \eq{homolat}]
\cite{JAV99,BUR02} and, as the lattice becomes deeper, it is
progressively more difficult to maintain the adiabaticity with
respect to these excitations. In deep lattices for the loading to
remain adiabatic it is therefore required that
\beq
\zeta(t)\equiv\left|{1\over J(t)}{\partial J(t)\over\partial
t}\right|\ll \omega_j(t) \,,\label{adiaba}
\eeq
for all the phonon mode energies $\omega_j$ during the turning up of
the lattice.

In Fig.~\ref{heightfig} we find the number squeezing to saturate
around $s$=20-30, indicating the point where an increasing number of
phonon modes is excited and the phonon mode frequencies no longer
satisfy the condition \eq{adiaba}. Consequently, the turning up of
the lattice potential is strongly nonadiabatic. Due to the
nonadiabaticity, the $s\ge 15$ cases exhibit significant excess
number fluctuations as compared to the ground state. After a short
time period over which $C_1$ remains constant, the large $\Delta
n_i$ evolve into large phase fluctuations and $C_1$ becomes
oscillatory and collapses.

As argued in Ref.~\cite{JAV99}, if the adiabaticity of a phonon mode
breaks down, the number fluctuations of the mode freeze to the value
that prevails at the time this occurs, i.e., when $\omega_j\sim
\zeta(t)$. If we naively apply this argument to the Bogoliubov
result in Eq.~\eqref{deltan} for the case where the adiabaticity
condition \eq{adiaba} breaks down for several modes with $\omega_j
\alt \zeta(t)$, we obtain
\beq
(\Delta n)^2\simeq \s_j \f{ \hbar\zeta_j(t_j)}{2UN_p}\,,
\eeq
where $\zeta_j(t_j)$ denotes the value of $\zeta(t)$ when we have
$\omega_j\sim \zeta(t)$ for the first time during the turning up of
the lattice for the $j$th phonon mode at time $t_j$. Since for all
$j$, $\zeta_j(t_j)$ is in the studied case roughly of the order of
$\omega$, we have the asymptotic value for $s\rightarrow\infty$,
\beq
(\Delta n_i)^2\sim {\hbar\omega\over U}\,,
\eeq
qualitatively similar to the results in Fig.~\ref{heightfig}.

We may also estimate in this case the time required to turn up the
optical lattice adiabatically from $s=10$ to $s=35$. If, for
simplicity, we assume that the optical lattice was ramped up in such
a way that $J(t)\simeq J_0 e^{-\kappa t}$, we obtain in \eq{adiaba}
a constant $\xi=\kappa$. In order to have an adiabatic loading of
the atoms into the lattice, we therefore require that $\kappa\ll
\omega_j$ for all the modes $j$. If we assume that the homogeneous
Bogoliubov result \eq{lowestphonon} for the lowest phonon mode
energy $\omega_{q,{\rm min}}$ still provides at least the correct
order of magnitude estimate at $s=35$ with the large filling factor
considered here, the adiabaticity condition is roughly satisfied if
$\kappa\ll 0.1/\omega$. The corresponding time required to have a
slow enough ramping from $s=10$ to $s=35$ is then $\Delta t \gg
40/\omega\simeq 0.6$s, where we have used the value of the trap
frequency $\omega=2\pi\times10$Hz (with the present set of
parameters this corresponds to the lattice spacing $d\simeq1.3\mu$m
for $^{87}$Rb). However, such a slow turning up process of possibly
tens of seconds in order to reach the Mott ground state is extremely
difficult to achieve experimentally, since the ground state lifetime
is limited by the losses due to the three-body collisions and the
spontaneous emission of photons.

A similar argument also applies to the atom number squeezing
experiment in Ref.~\cite{ORZ01}. In Ref.~\cite{ORZ01} a BEC was
initially confined in a harmonic trap. The atom cloud was then split
by continuously turning up an optical lattice and the atoms were
interfered at different final heights of the lattice potential. The
atom number squeezing in individual lattice sites was calculated
from the phase fluctuations by measuring the loss of visibility in
the interference fringes. The lattice had about 12 sites with
$\sim$1000 atoms in the central well. The lattice depths of up to
$s=50$ were reached, corresponding to $n_iU/J\sim10^5$. Here $n_iU$
was estimated from the size of the atom cloud and it varied
considerably during the ramping, since the lattice light field also
changed the transverse confinement depending on the lattice height.
If we use the same simplified analysis as before to estimate the
time required to turn up the lattice adiabatically from $s=10$ to
$s=50$, we obtain $\Delta t\gg 50$ms. Consequently, it is not
surprising that the ramping-up time 200ms used in the experiments
was not sufficiently slow to reach the ground state of the atoms.
The adiabaticity condition $\Delta t\gg 50$ms is not as severe as in
our simulation example, but in the system of Ref.~\cite{ORZ01} even
the lattice height $s=50$ is still far below the Mott transition
point.

As we already argued in Ref.~\cite{ISE05}, the requirement of a very
slow ramping-up time generally makes it difficult to reach the Mott
insulator ground state experimentally when there are many atoms per
lattice site for a large number of sites. This indicates that
producing technologically interesting strong atom number squeezing
with large occupation numbers has serious limitations. In the
lattices with small filling factors, that have been so far used in
the Mott insulator state experiments, the above argument is no
longer valid due to rapidly emerging energy gap and much larger
quantum fluctuations even in a shallow lattice.

We have shown that for large filling factors the squeezing of atom
number fluctuations within individual lattice sites, resulting from
the turning up of the lattice potential, saturates for deep
lattices. The saturation of the number squeezing for strong lattices
was experimentally observed in Ref.~\cite{ORZ01}. As we already
pointed out, such a system is not tightly elongated, but we can
still make qualitative comparisons to the experimental data.
Although the saturation was assumed in Ref.~\cite{ORZ01} to be an
artifact of the analysis method of the interference measurement, we
also numerically find similar saturation effect in a qualitative
agreement with the experiment, without including the effects of the
interference experiment. As we argued here, such saturation may
result as a consequence of the nonadiabaticity of the loading
process.

If the loading is sufficiently rapid or the final lattice
sufficiently high, so that the adiabaticity breaks down for a large
number of modes, the optimal number squeezing is proportional to the
ramping speed itself and the nonlinearity. Both in
Fig.~\ref{heightfig} and in Ref.~\cite{ORZ01} the squeezing
saturates at about 15dB when $n_iU/J\sim10^4$. The ramping-up time
$\tau\simeq 4000\hbar/E_r$ in Ref.~\cite{ORZ01} is one order of
magnitude longer than in Fig.~\ref{heightfig}, but this is
compensated by the weaker hopping amplitude $J$, so that the
saturation roughly occurs at the same value of $\omega_n\tau$.

Due to the nonadiabaticity of the loading of atoms into the lattice,
the system in Ref.~\cite{ORZ01} may not have been in its ground
state during the turning up of the lattice potential. Similarly to
our numerical example, the nonadiabaticity in the experiment can
induce larger phase fluctuations than those in the Heisenberg
minimum uncertainty state for the atom number and phase.
Additionally, an inhomogeneous phase profile can reduce the
visibility of interference fringes similarly to phase fluctuations.
Consequently, not all the experimentally detected phase noise in the
loss of interference fringes may directly correspond to the atom
number squeezing. Although the increased phase noise in the
experiment therefore did not provide an entirely conclusive measure
of the atom number fluctuations, our numerical simulations,
nevertheless, seem to indicate that a considerable number squeezing
may have been present also in Ref.~\cite{ORZ01}.

The effect of increased phase fluctuations due to a rapid turning up
of the lattice may even be more pronounced in the absence of strong
radial confinement, as in Ref.~\cite{ORZ01}, as a result of the
nonlinear coupling between the radial and the axial modes. The
numerical solution of GPE in 1D, in Ref.~\cite{BAN02}, and in 3D, in
Ref.~\cite{MCK05}, was compared to the experimentally observed phase
noise using the parameters of Ref.~\cite{ORZ01}. Although this model
cannot incorporate any quantum effects, the numerics showed signs of
a reduced interference visibility due to inhomogeneous phase profile
\cite{PLA04}.

\subsubsection{The effect of initial temperature and nonlinearity}\label{initemp}

\begin{figure}
\includegraphics[width=0.49\columnwidth]{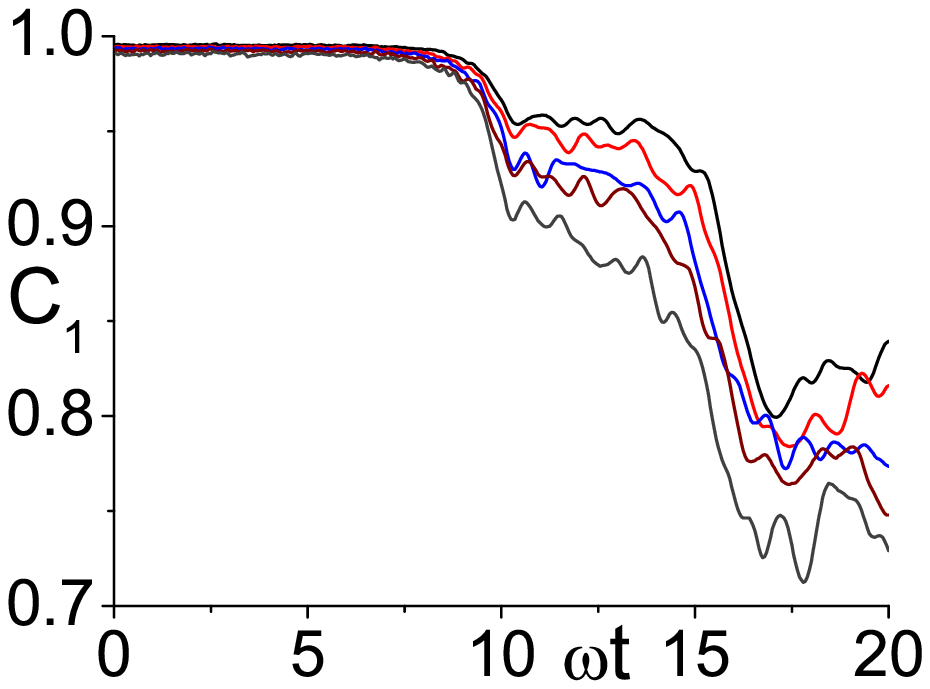}
\includegraphics[width=0.49\columnwidth]{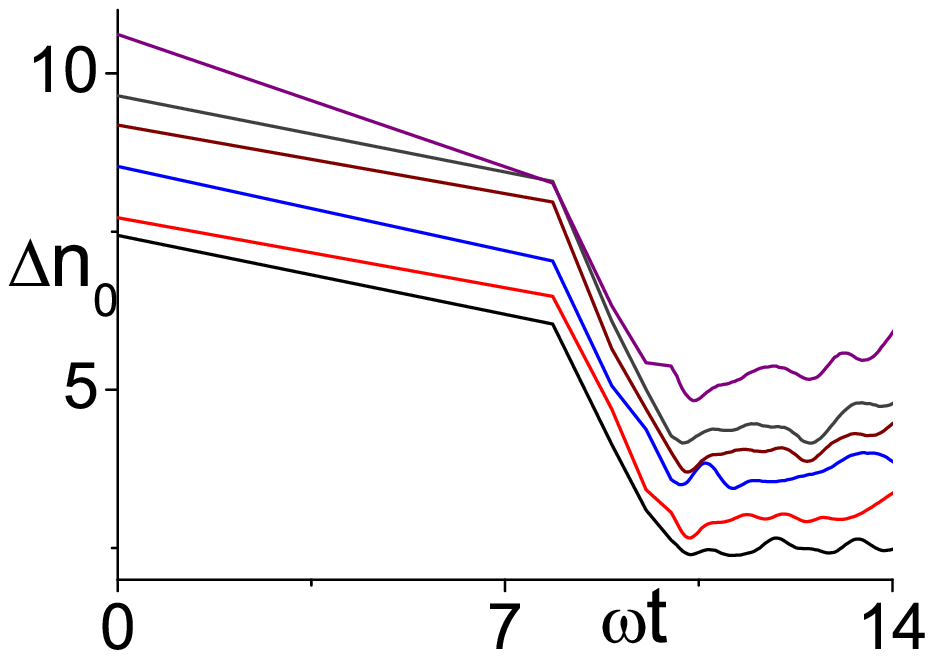} 
\includegraphics[width=0.49\columnwidth]{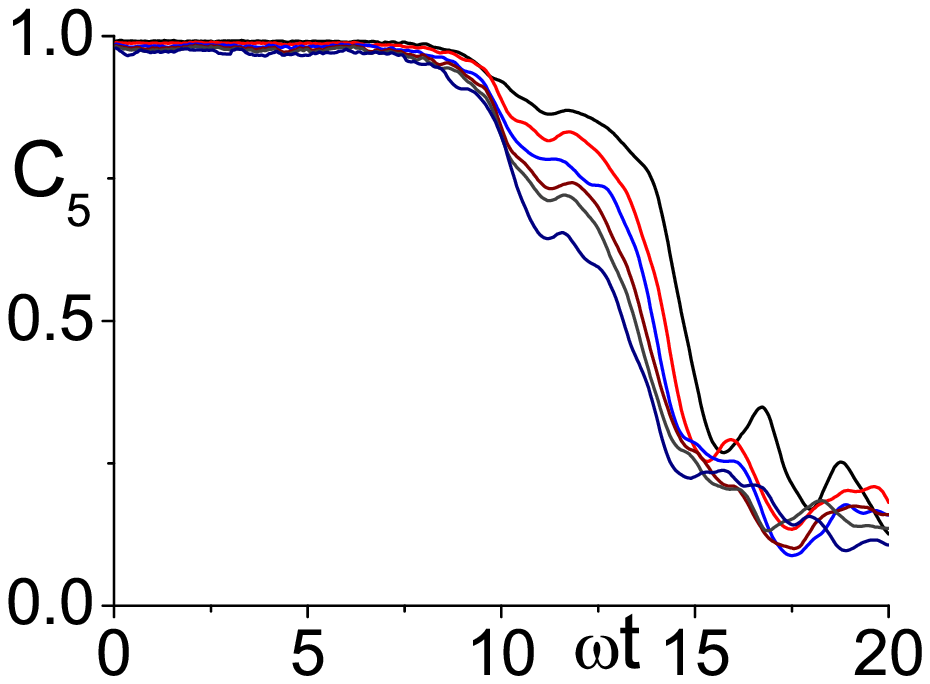}
\includegraphics[width=0.49\columnwidth]{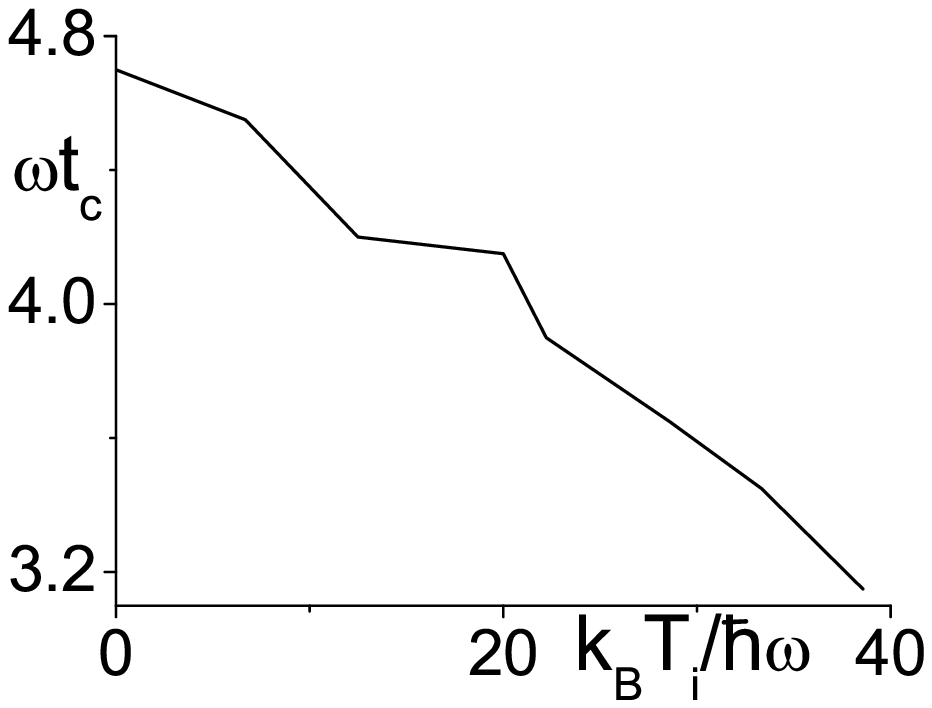}
\vspace{-0.4cm} \caption{The phase coherences $C_1$ (top left),
$C_5$ (bottom left) and the number fluctuations $\Delta n_0$ (top
right) for initial temperatures (curves from top)
$k_BT_i/\hbar\omega=0,12.5,22.2,33.3,38.5$ ($C_5$ also with 28.5).
The phase collapse time $t_c$ (bottom right) is evaluated at
$C_5=0.5$, subtracting the ramping-up time. Here
$N_0g_{1D}=100\hbar\omega l$, $N_0=2000$, $\omega\tau=10$, and
$s=20$.} \label{T-fig}
\end{figure}

Finite initial temperature increases the initial noise in the Wigner
distribution [Eq.~\eqref{wigner}] of the TWA as a consequence of the
finite temperature noncondensate fraction in the excited levels.
This is expected to affect the coherence between the different sites
(as we saw in Section~\ref{toy-system}) and the fluctuations in the
atom number $\Delta n_0$. In Fig.~\ref{T-fig} we show $\Delta n_0$
and the phase coherence $C_1$ for different initial temperatures
$T_i$ for final lattice height $s(\tau)=20$. Here $(\Delta n_0)^2$
increases exponentially as a function of $T_i$. The phase coherence
$C_5$ between the central well and its fifth neighbor decays
significantly faster than $C_1$. The dependence of the phase
collapse time $\tau_c$ on the initial temperature $T_i$ is
approximately linear (compare to Fig.~\ref{firstnumerics}). At
$s=20$ the effects of the harmonic trap are already significant in
$C_5$, since the potential energy difference due to the harmonic
trap between the central lattice site and its fifth nearest neighbor
exceeds the tunneling energy
$V_h(j=5)-V_h(j=0)\simeq2\hbar\omega\agt n_0J$.

\begin{figure}
\includegraphics[width=0.49\columnwidth]{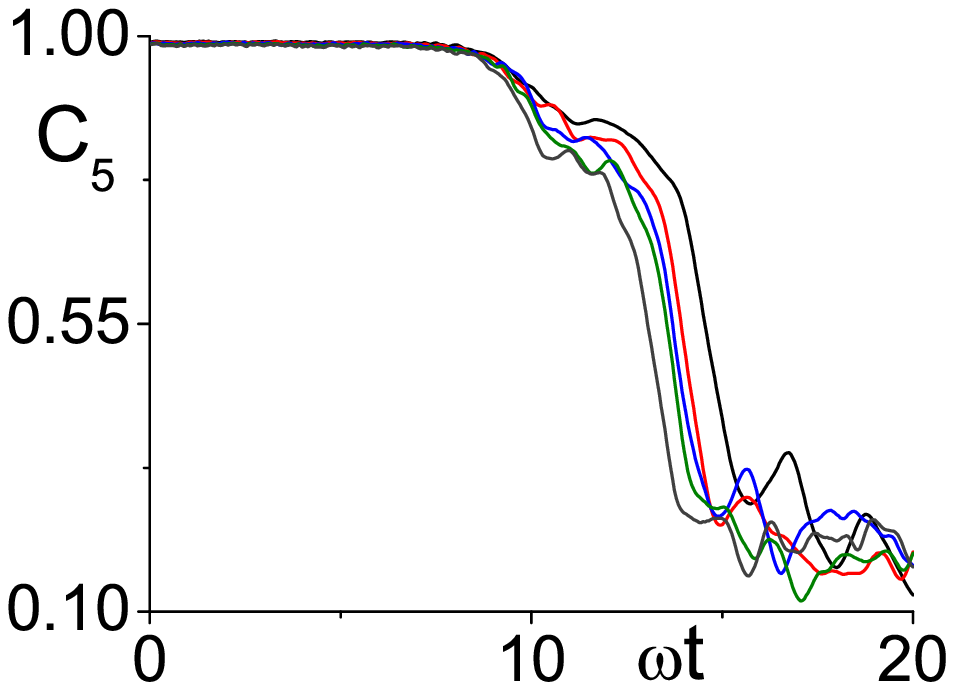}
\includegraphics[width=0.49\columnwidth]{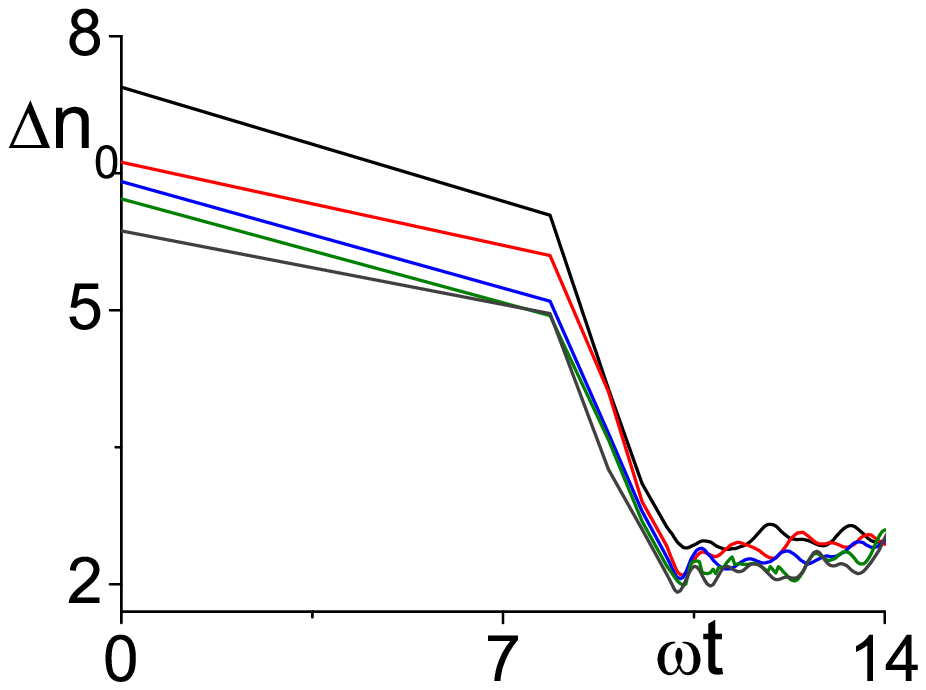}
\includegraphics[width=0.45\columnwidth]{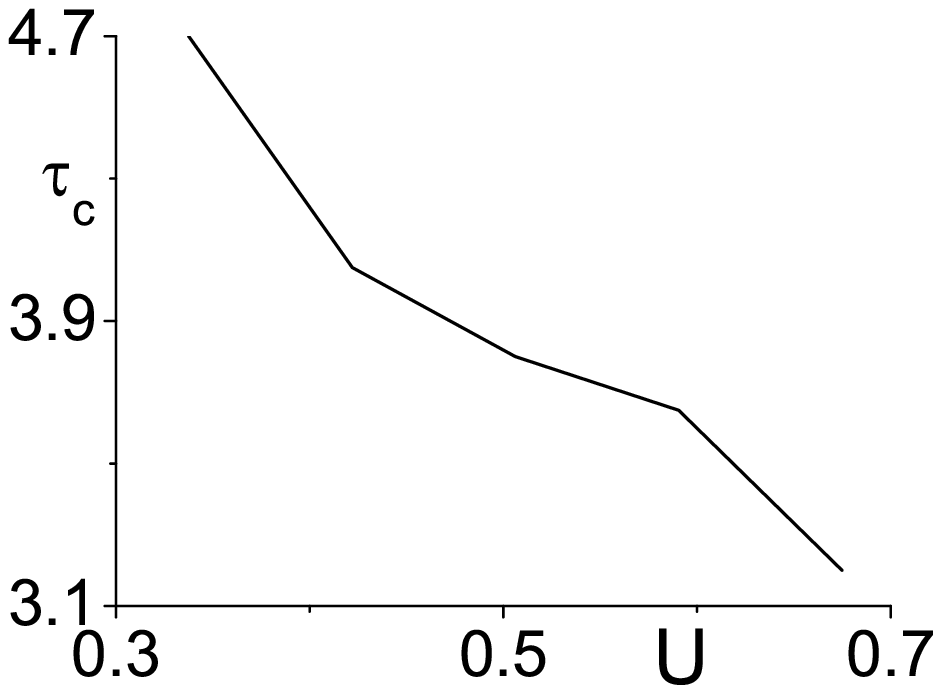}
\includegraphics[width=0.53\columnwidth]{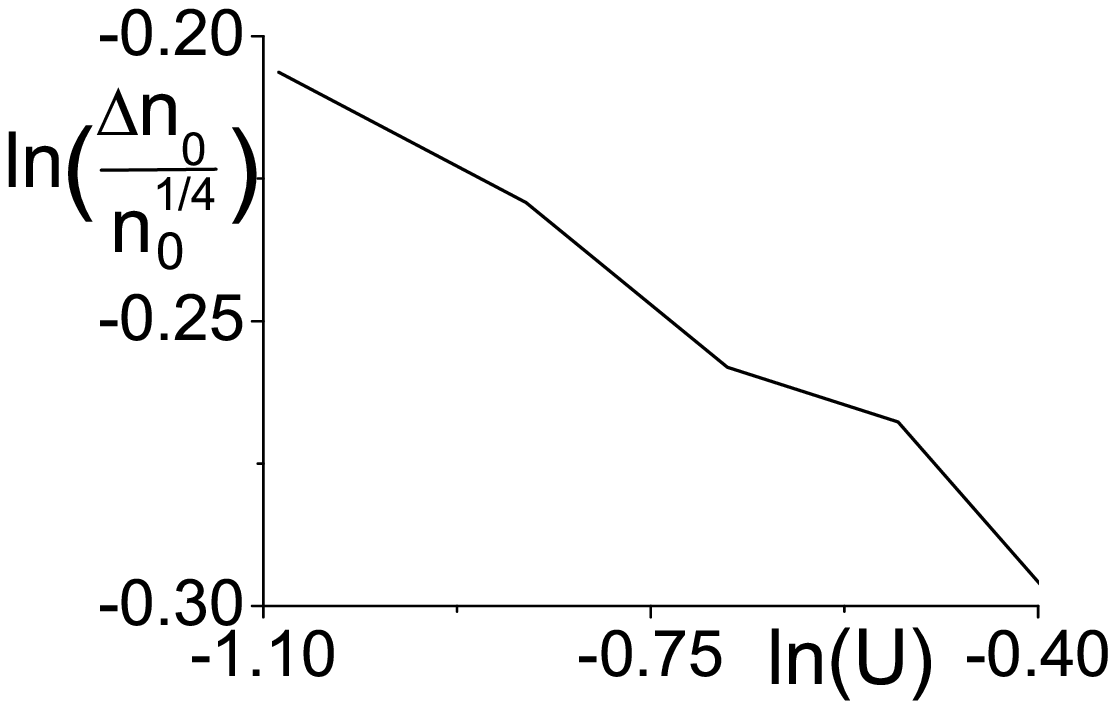}
\vspace{-0.4cm} \caption{The coherence function $C_5$ (top left) and
the number fluctuations in the central well (top right). The curves
from top represent $N_0g_{1D}/(\hbar\omega l)=100, 125, 150, 175,
200$. The collapse time (evaluated when $C_5=0.5$ and subtracting
the ramping-up time) (bottom left) and $\Delta n_0/n_0^{1/4}$
(bottom right) as a function of the on-site interaction $U$ (in the
units of $\hbar\omega$) at $T_i=0$. Here $N_0=2000$, $s=20$, and
$\omega\tau=10$.} \label{nonlin-fig}
\end{figure}

In order to estimate the effect of the nonlinearity we varied
$N_0g_{1D}/(\hbar\omega l)$ from $100$ to $200$ for the case of the
final lattice height $s=20$, the ramping-up time $\omega\tau=10$,
and the initial temperature $T_i=0$. In Fig.~\ref{nonlin-fig} we
show the results for the phase coherence $C_5$ and the atom number
fluctuations $\Delta n_0$ in the central well. The dependence of the
phase collapse time on the nonlinearity is obtained by choosing the
time when $C_5=0.5$. We compare the numerical TWA results for
$\Delta n_0$ to the analytic tight-binding result in the ground
state of the uniform lattice \eqref{deltan-T0} by fitting the TWA
results according to $(\Delta n_0)/\sqrt{n_0}\propto U^c$. Here the
exponent $c$ of the on-site nonlinearity is the fitting parameter.
The TWA simulations yield $c\simeq-0.3$, as compared to the ground
state result $c=-0.5$ of \eq{deltan-T0}. In Fig.~\ref{nonlin-fig},
the $(\Delta n_0)/\sqrt{n_0}$ plot was obtained by averaging over
the time interval $4/\omega$ after the lattice was turned up.

\subsubsection{Validity of the classical GP theory}\label{valGPE}

The classical mean-field theories, in particular the GPE, have been
very successful in describing the full multi-mode dynamics of
weakly-interacting, harmonically trapped atomic BECs. However, the
GPE has severe limitations in optical lattices where the
interactions are enhanced, since the classical GPE disregards
thermal and quantum fluctuations, decoherence, and the information
about quantum statistics. It is especially interesting to study the
limits of validity of the GPE in shallow optical lattices. In
Ref.~\cite{RUO05} it was shown that the experimentally observed
damping of the dipolar motion of atoms in a very shallow lattice
with $s=0.25$ \cite{FER05} resulted from the large ratio $g_{1D}/N$.
If the atom number in the simulations was increased, or $g_{1D}/N$
decreased, while at the same time keeping the chemical potential
$\propto Ng_{1D}$ fixed, the damping rate was reduced exponentially
and the results approached the classical GPE limit. Here we study
the turning up of the lattice for much larger filling factors and,
hence, smaller $g_{1D}/N$.

\begin{figure}
\includegraphics[width=0.47\columnwidth]{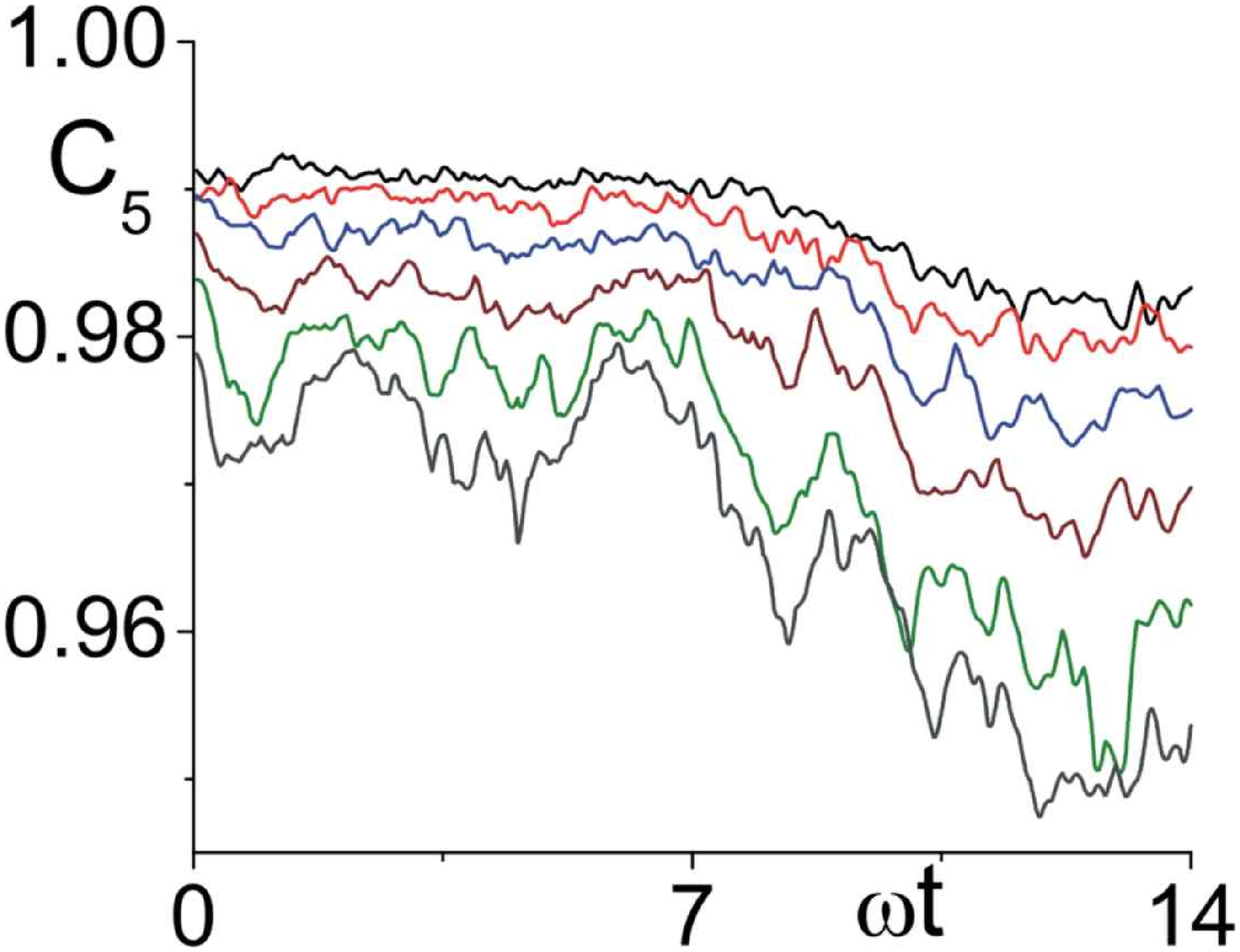}
\includegraphics[width=0.47\columnwidth]{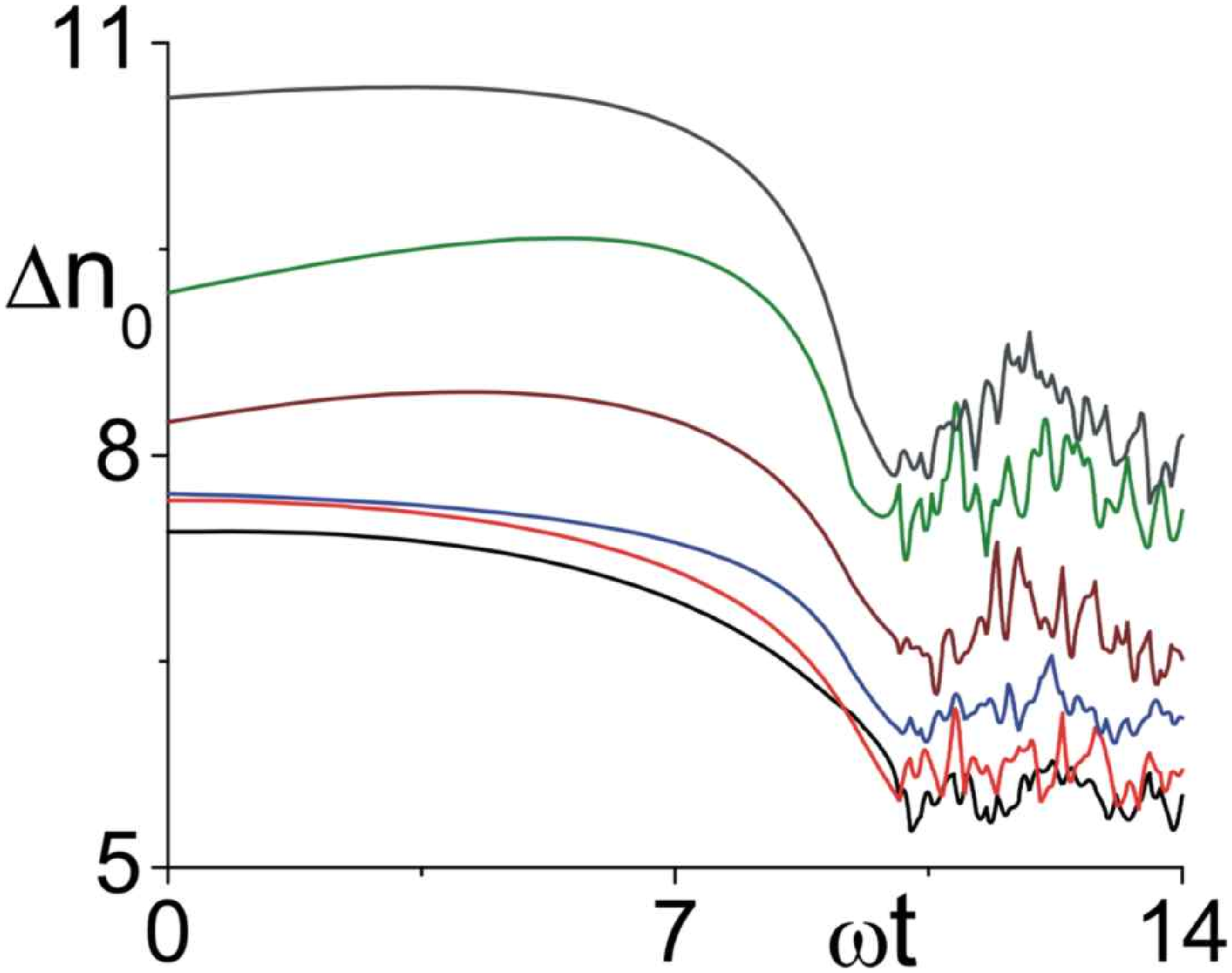}
\includegraphics[width=0.49\columnwidth]{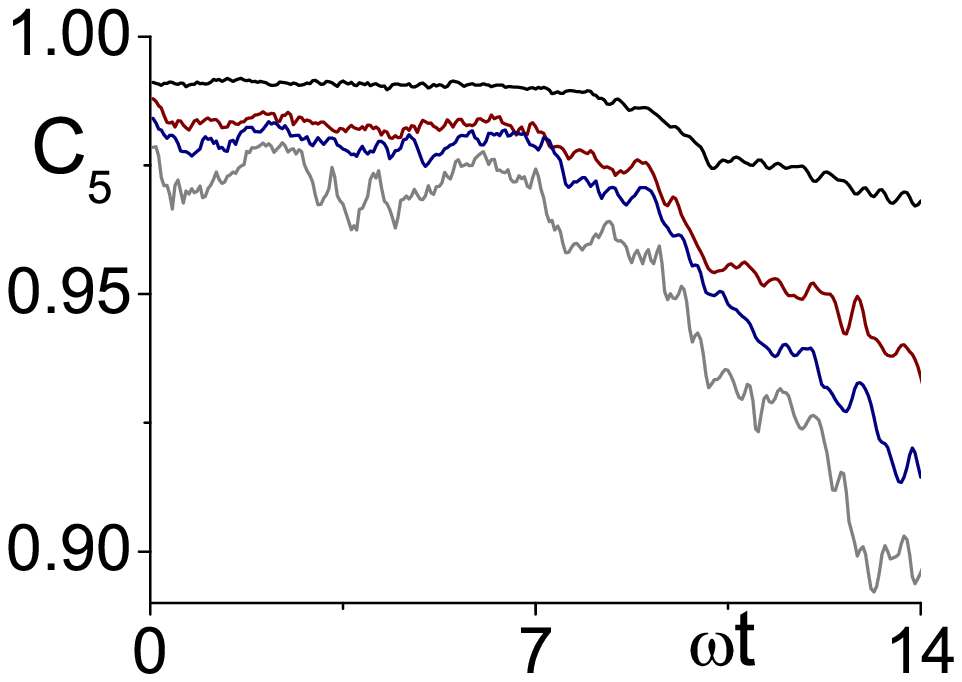}
\includegraphics[width=0.49\columnwidth]{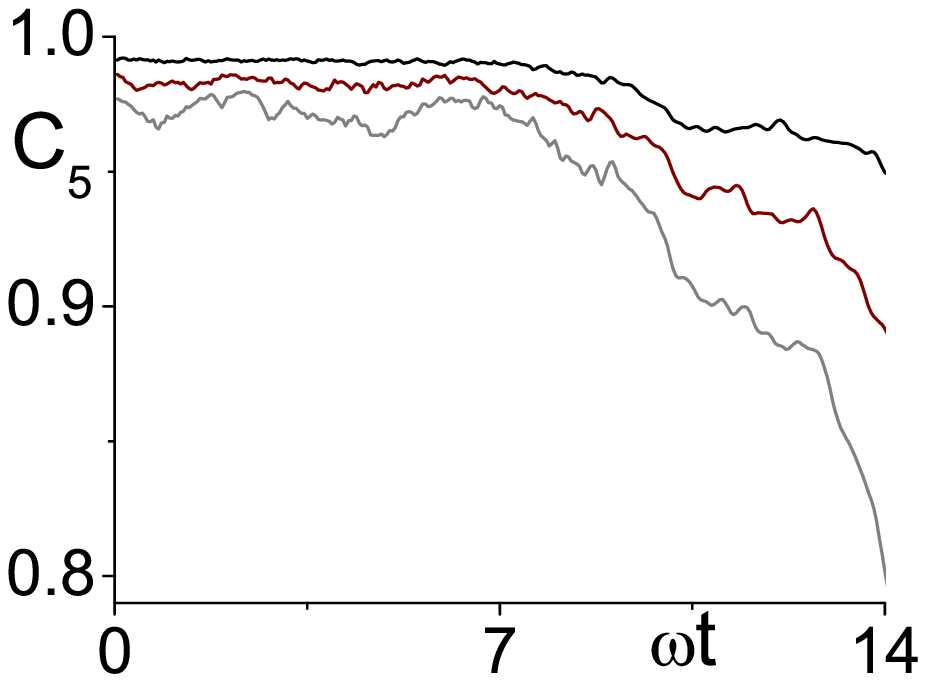}
\vspace{-0.4cm} \caption{The coherence $C_5$ between the central
well and the fifth neighbor (top left) and the atom number
fluctuations in the central well $\Delta n_0$ (top right) for the
case of the final lattice height $s=5$. The curves from top
represent the initial temperatures
$k_BT_i/\hbar\omega=0,6.67,12.5,22.2,33.3,38.5$. The phase coherence
$C_5$ for $s=8$ (bottom left) and for $s=10$ (bottom right). The
curves from top represent the initial temperatures
$k_BT_i/\hbar\omega=0,22.2,28.6,38.5$ and $0,22.2,38.5$,
respectively. The nonlinearity $N_0g_{1D}=100\hbar\omega l$,
$N_0=2000$, and the ramping-up time $\omega\tau=10$.}
\label{s5-vs-Tini}
\end{figure}

In Fig.~\ref{s5-vs-Tini} we show the coherence between the central
well and its fifth neighbor for different final lattice heights and
different temperatures and the atom number fluctuations in the
central well for $s=5$. The number of atoms in the central site
$n_0\simeq$ 90-100. For the case of $s=5$ the phase coherence
remains high at low initial temperatures. The number fluctuations at
$T_i=0$ are weakly sub-Poissonian $\Delta n_0\simeq 5.5<10$ and are
increased due to the initial finite temperature. However, for $s=8$
and $s=10$, the TWA results are notably different from the classical
GPE dynamics, even at zero temperature. For $s=8$ and $s=10$ the
loss of coherence at $T_i=0$ due to vacuum fluctuations is already
comparable to the loss of coherence at $s=5$ due to thermal
fluctuations at $k_B T_i=38.5\hbar\omega$. (When the initial thermal
population is close to ten percent.) For $s=10$, the atom number
fluctuations are also more sub-Poissonian $\Delta n_0\simeq 3.9<10$
at $T_i=0$; see Fig.~\ref{heightfig}.

\subsubsection{Change of temperature during the splitting}\label{ret-lat}

\begin{figure}
\includegraphics[width=0.49\columnwidth]{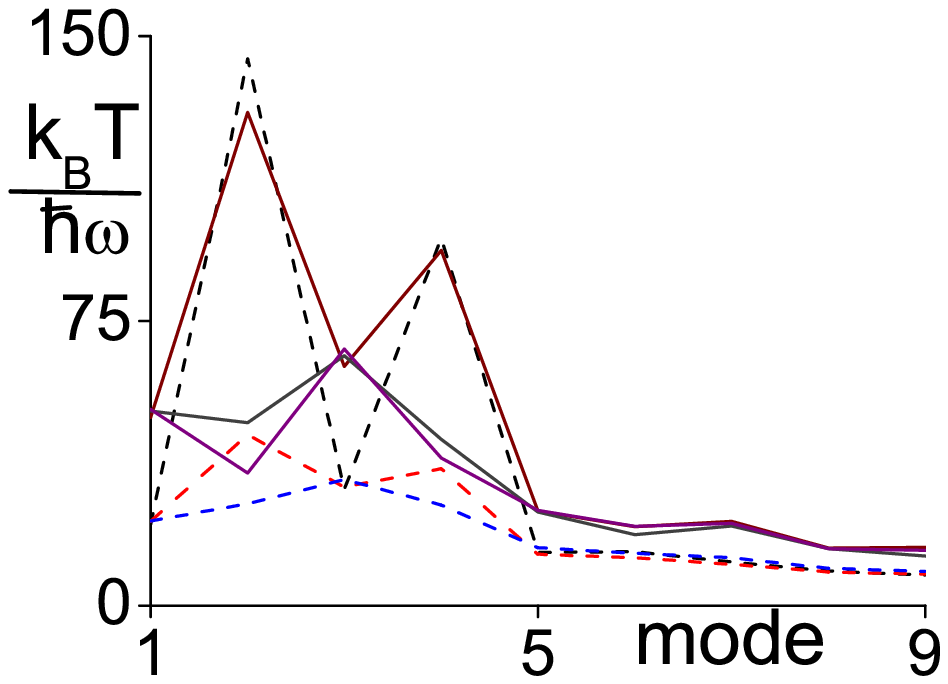}
\includegraphics[width=0.49\columnwidth]{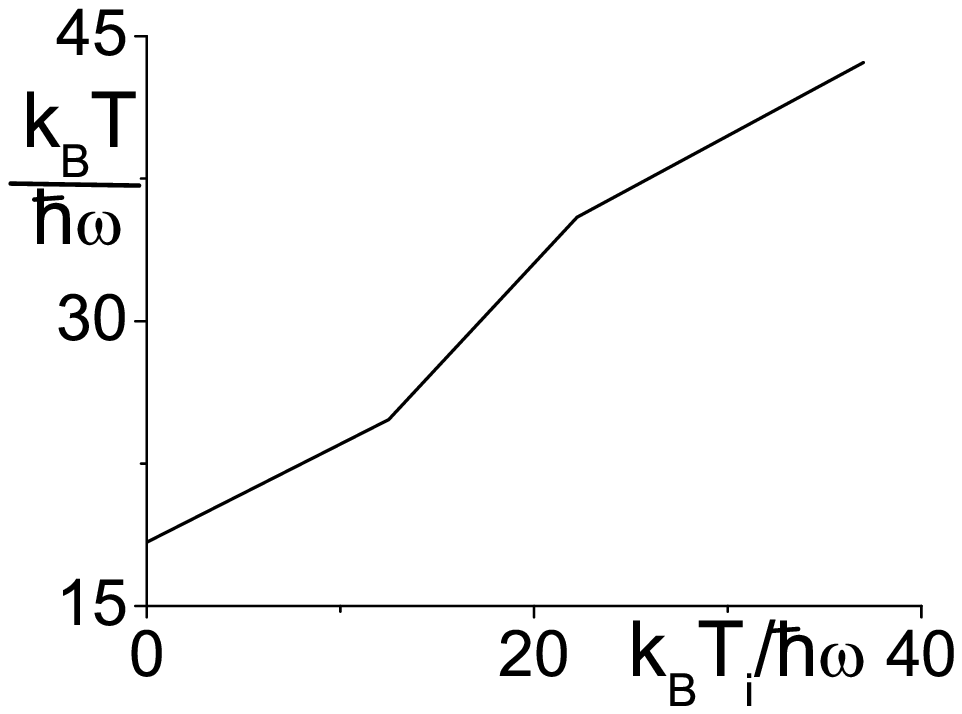}
\vspace{-0.4cm} \caption{The contribution of the lowest modes to
temperature (left) for $k_BT_i/\hbar\omega=12.5$ (dashed line) and
37 (solid line). Curves from top to bottom represent
$\omega\tau=10,20,30$ in both cases. The average temperature of the
first five modes after the ramping $\omega\tau=30$ (right) for an
initial temperature $k_BT_i=0,12.5,22.2,37$. Here
$g_{1D}=0.015\hbar\omega l$, $N_0=2000$, and $s=5$.}
\label{rethermalizationfig}
\end{figure}
The optical potential also affects the temperature of the atom
vapor. If the lattice potential is turned up adiabatically, the
population of each mode remains constant and temperature $T$ can
change dramatically, as the contribution of each mode to $T$ changes
by the ratio of the final and initial mode energies
$\omega_j^{(f)}/\omega_j^{(i)}$. An adiabatic increase in the
lattice strength may both increase or lower $T$, depending on
whether the excited band is occupied \cite{BLA04} and in the
experiments the condensation temperature has been found to be
sensitive to the lattice height \cite{BURG02}. In
Fig.~\ref{rethermalizationfig} we estimated the `temperature' of
several lowest phonon modes in the TWA simulations by evaluating the
corresponding occupation numbers $\overline n_k$. This was obtained
by calculating the projection of $\psi_W$ to the Bogoliubov modes of
the BHH using \eq{chiexpr}, as explained in
Appendix~\ref{bogo-approach}. The averages are taken over a time
period before any significant rethermalization occurs after the
ramping \cite{rethcom}. The modes 2 and 4 are highly excited for the
case of short $\tau$, due to the nonadiabatic loading. The
excitations are damped out at higher initial temperatures $T_i$ and
for the case of the slowest ramping $\omega\tau=30$, representing
the situation where $\omega_{2}\tau\gg 1$ and $\omega_{4}\tau\gg1$.
It is interesting to note that the excitations of the forth mode are
only damped out when the rate of change in the tunneling amplitude
$\zeta$ is much smaller than the corresponding mode energy, or when
$\omega_4\simeq26\zeta(\tau)$. This is more restrictive condition
than the one found in Ref.~\cite{JAV99}. For very fast ramping
$\omega\tau\alt3$, the variation of $T_i$ is already completely
dominated by the excitations due to the rapid turning up of the
lattice.

\subsubsection{Turning up the nonlinearity} \label{ramp-non-lin}

The numerical solutions for the initial equilibrium state in the TWA
simulations may in some cases, especially in higher dimensions, be
difficult to obtain. In Ref.~\cite{POL03c} the ideal, noninteracting
condensate was used as an initial state for the TWA simulations, but
the interaction constant was first continuously ramped from zero up
to some final value, before the actual atom dynamics was studied. By
means of turning the nonlinearity up slowly enough, the goal was to
produce the initial state of an interacting system. If the ramping
is slow enough, the dynamics is then expected to resemble to that of
the interacting system. Here we study the atom dynamics by first
linearly turning up the interaction constant from $g_{1D}^{\rm
initial}=0$ to a final value $g_{1D}^{\rm final}$ during the time
$\tau_g$, before starting to ramp up the optical potential.
Experimentally, such a technique could possibly be employed by means
of using Feshbach resonances to tune the value of the scattering
length $a$.

The field operator for the initial state of the noninteracting
system is that of the ideal harmonic oscillators in thermal
equilibrium:
\begin{align} \label{initial-non-interacting}
\h\psi(x,t=0)=& \f{\h\alpha_0}{\pi^{1/4}l^{1/2}}\exp\(-\f{x^2}{2l^2}\)\nonumber \\
&+ \sum_{j>0}\h\alpha_jR_j H_j({x/l})\exp\(-\f{x^2}{2l^2}\)\,,
\end{align}
where we have explicitly separated the BEC mode, $H_j$ is the $j$th
Hermite polynomial, and $R_j\equiv (\sqrt{\pi}2^j j!l)^{-1/2}$. The
sampling of the Wigner distribution for $\alpha_j$ and $\alpha_0$ to
generate $\psi_W(x,t=0)$ is then performed exactly as in the
interacting case.

\begin{figure}
\includegraphics[width=0.49\columnwidth]{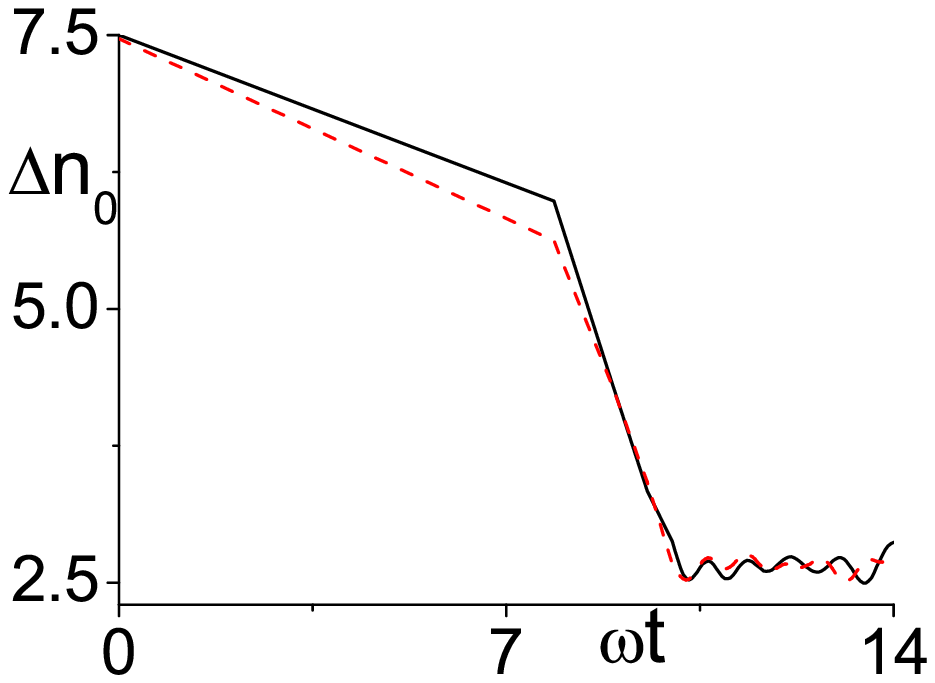}
\includegraphics[width=0.49\columnwidth]{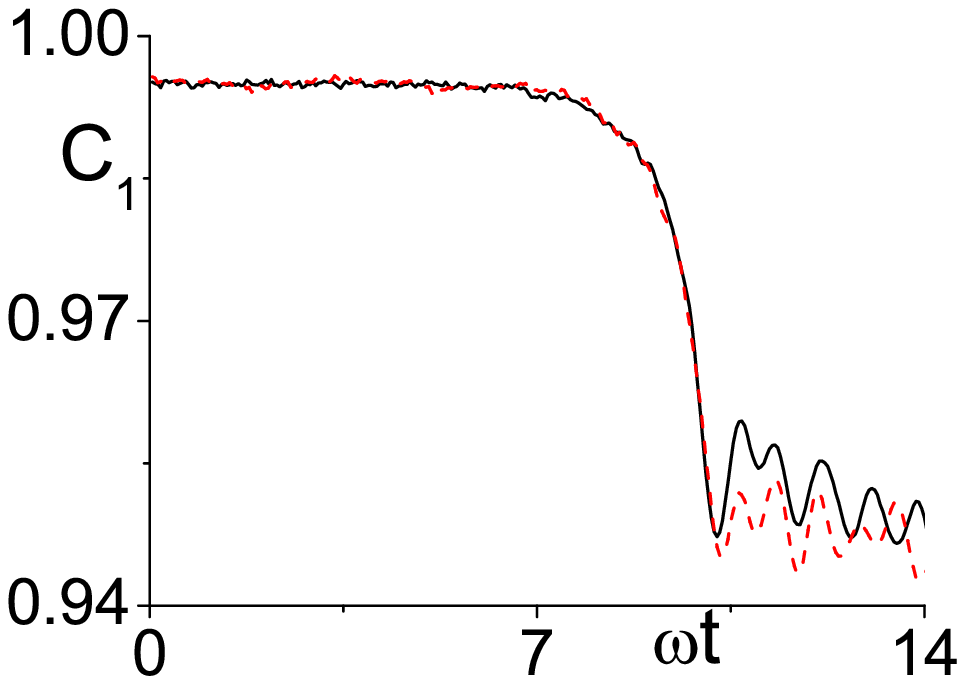}
\includegraphics[width=0.49\columnwidth]{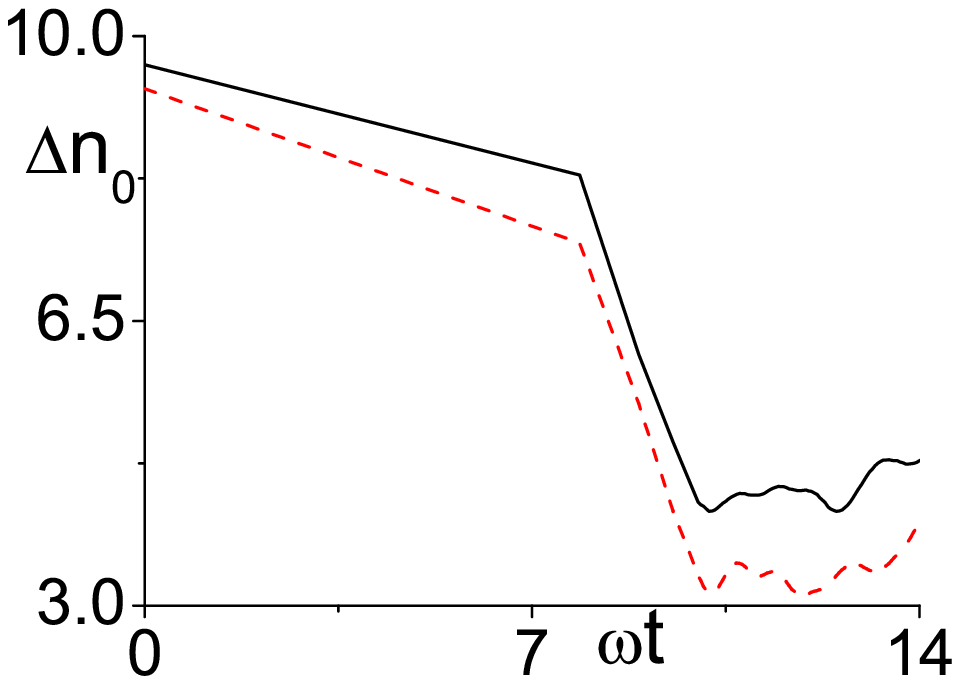}
\includegraphics[width=0.49\columnwidth]{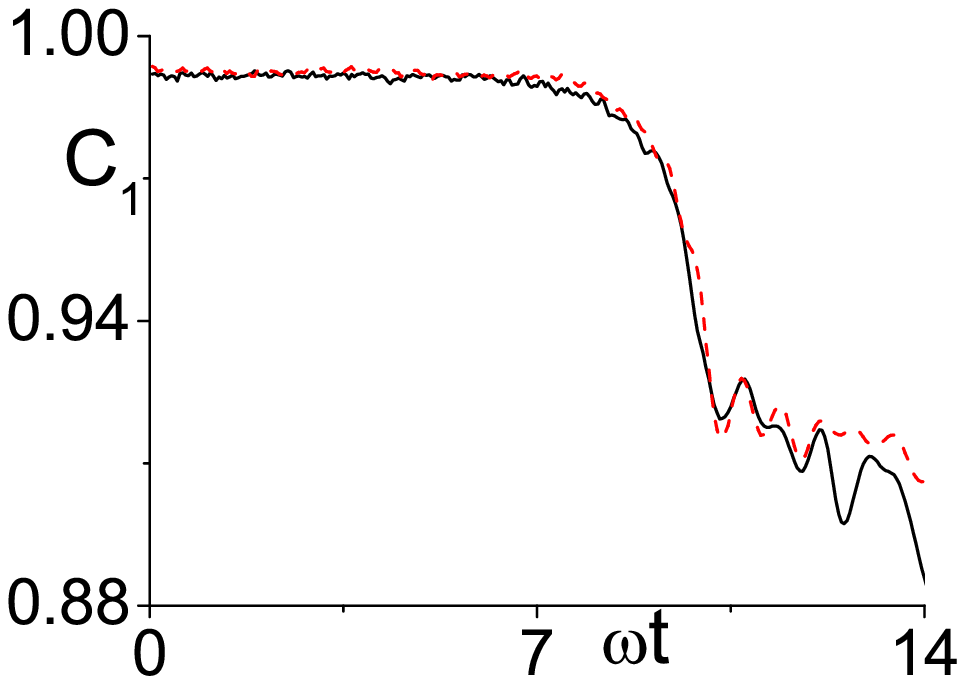}
\vspace{-0.4cm}\caption{The phase coherence and the number
fluctuations for an initially noninteracting system ($g_{1D}^{\rm
initial}=~0$) with the interactions linearly turned up to the value
$g_{1D}^{\rm final}=g_{1D}=0.05\hbar\omega l$ (dashed red line) and
in an initially interacting system (solid black line), obtained by
solving the Bogoliubov equations for $g_{1D}=0.05\hbar\omega l$ with
$N_0=2000$, $s=20$, and $\omega\tau=10$. The initial temperature is
$k_BT_i/\hbar\omega=6.67$ (on top) and 33.33 (at bottom) and the
ramping-up time of the nonlinearity $\omega\tau_g=15$. The results
for the evolution of the ideal gas describe its dynamics after the
end of the ramping of the nonlinearity. }\label{rampingfig}
\end{figure}

In Fig.~\ref{rampingfig} we show the TWA simulation results for the
initially noninteracting system for which we first slowly turn up
the interactions to the value $g_{1D}=0.05\hbar\omega l$, before
ramping up the optical lattice potential. We also show the results
for the initially interacting system for which the initial state is
obtained by solving the Bogoliubov equations with
$g_{1D}=0.05\hbar\omega l$. The initial temperature of the
noninteracting case is set to be equal to the temperature of the
interacting system we want to compare. Therefore the condensate and
the noncondensate populations in the initial state of the TWA
simulations are slightly different in the interacting and
noninteracting cases.

In the example case the two approaches yield surprisingly similar
results at low $T_i$. At higher $T_i$ the atom number fluctuations
in the case of the ideal gas initial state are smaller. This is most
likely due to the lack of any notable rethermalization in the 1D TWA
simulations \cite{rethcom}, so the turning up of the nonlinearity
fails to produce the thermal equilibrium state.

\subsubsection{Sampling of the noise for the initial ground state mode}\label{grnoise}

\begin{figure}
\includegraphics[width=0.49\columnwidth]{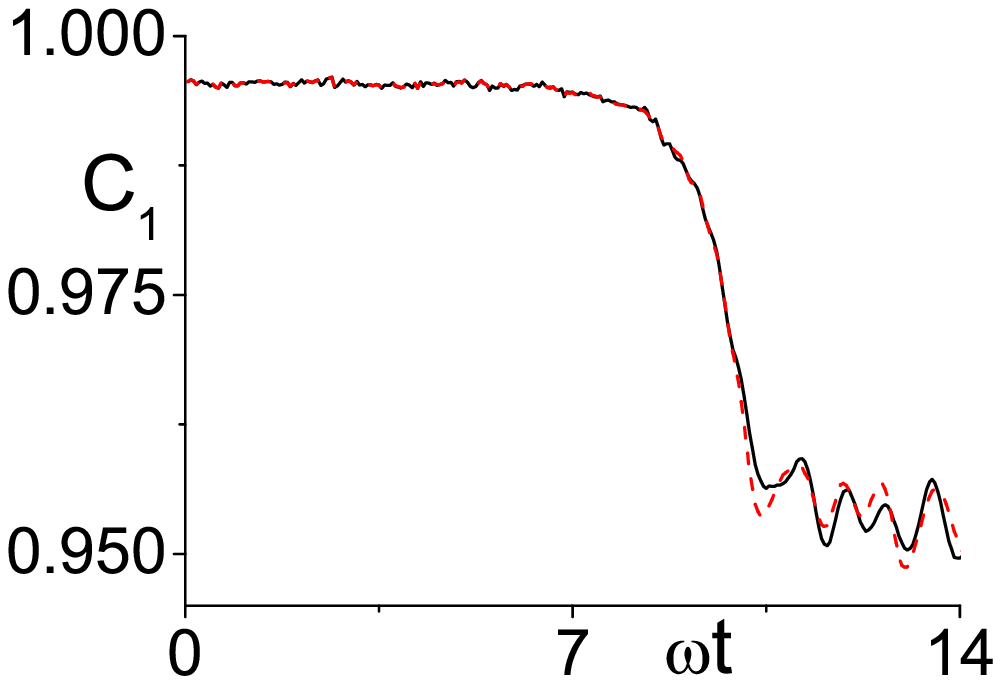}
\includegraphics[width=0.49\columnwidth]{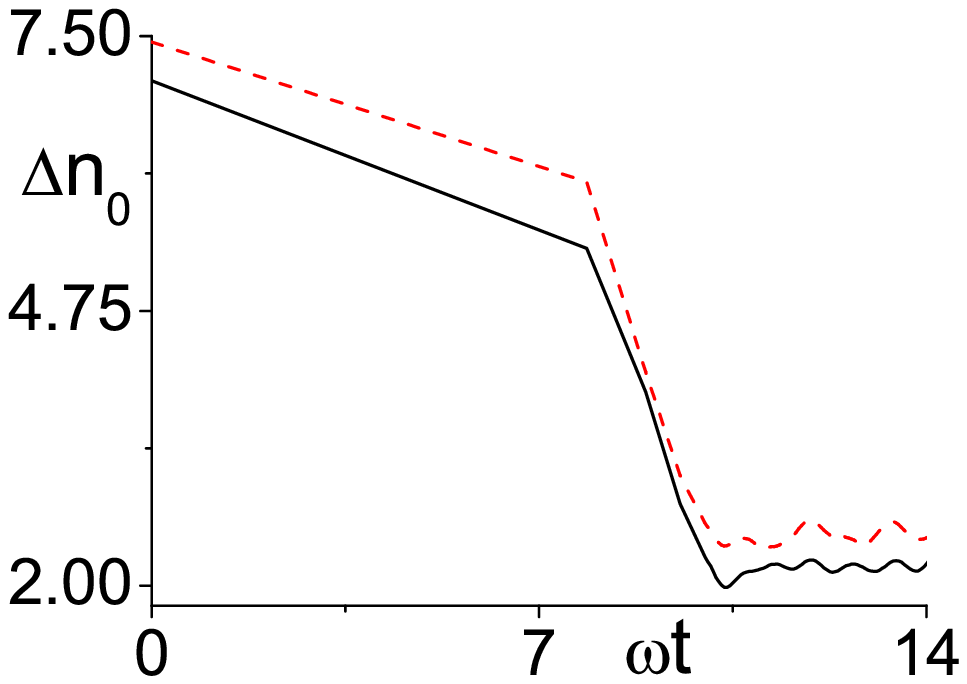}
\vspace{-0.4cm} \caption{The phase coherence $C_1$ and the atom
number fluctuations $\Delta n_0$ for cases where the initial ground
state was simulated classically with a fixed number of atoms (solid
black line) and where the initial ground state included quantum
fluctuations while assuming a coherent state (dashed red line). Here
$T_i=0$, $N_0g_{1D}=100\hbar\omega l$, $N_0=2000$, $s=20$, and
$\omega\tau=10$. }\label{sampling}
\end{figure}

In the TWA simulations the quantum fluctuations of the initial BEC
mode are included by assuming the BEC to be in a coherent state, so
that the initial state is sampled according to the corresponding
Wigner distribution in \eq{wc}. Although the condensate mode is
expected to exhibit sub-Poisson atom number fluctuations, the
advantage of the coherent state representation is that the Wigner
distribution is positive. For instance, the Wigner distribution of a
number state with $n$ atoms is nonpositive:
$W(\alpha,\alpha^*)=2(-1)^n\exp{(-2|\alpha|^2)}L_{n}(4|\alpha|^2)/\pi$
\cite{GAR}, where $L_n$ denotes the Laguerre polynomial, and cannot
be represented in the TWA simulations without doubling the dimension
of the phase space \cite{PLI01} analogously to the positive P
representation \cite{walls}. However, as we already noted earlier,
the particular representation is not so crucial, since the main
contribution to the matter wave coherence is due to the thermal and
quantum fluctuations of low-energy phonons and the quantum
fluctuations of the initial state of the BEC mode are not very
important. In Fig.~\ref{sampling} we show the TWA results for the
typical physical parameters of our system both when the BEC mode is
sampled according to \eq{wc} (with Poisson atom number fluctuations)
and when we entirely ignore the quantum fluctuations of the BEC mode
and treat it classically having a fixed number of atoms. A classical
treatment of the ground state does not affect the prediction for the
phase coherence along the lattice, but produces slightly smaller
atom number fluctuations in the lattice sites.

\subsubsection{Importance of the symmetric ordering}\label{ordering}

The TWA simulations return quantum expectations values for the
operators that are symmetrically ordered. This generally creates
significant complications in interpreting the simulation results in
terms of normally ordered expectation values, as we already
emphasized in Section \ref{interest}. However, as we here
demonstrate, the correct transformation between the different
operator orderings is very crucial in obtaining the correct physical
results.

Since the P-representation of the density matrix returns the
expectation values of normally ordered operators, e.g.,
$\<\alpha_j^*\alpha_j \>_P=\< \h\alpha_j^\d\h\alpha_j \>$, we may
also call the stochastic dynamics that assumes the operator
expectation values to be normally ordered as the ``truncated P
approximation" (TPA) \cite{SIN02}. The P-representation is singular
at $T=0$, but is often used to study finite temperature systems. A
review of its mathematical properties may be found in
Refs.~\cite{walls,GAR}.

\begin{figure}
\includegraphics[width=0.49\columnwidth]{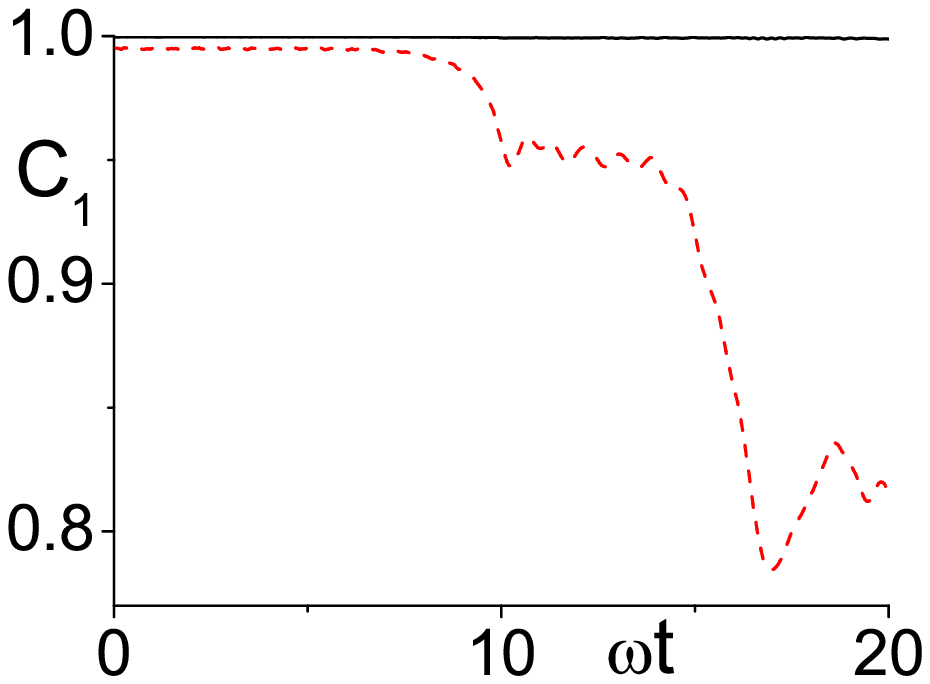}
\includegraphics[width=0.49\columnwidth]{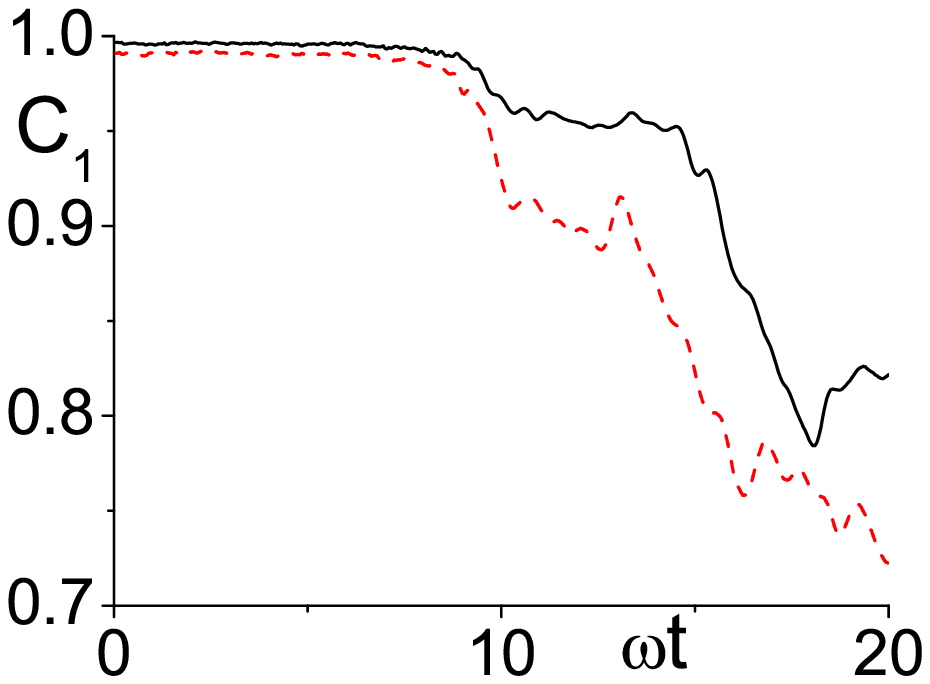}
\includegraphics[width=0.49\columnwidth]{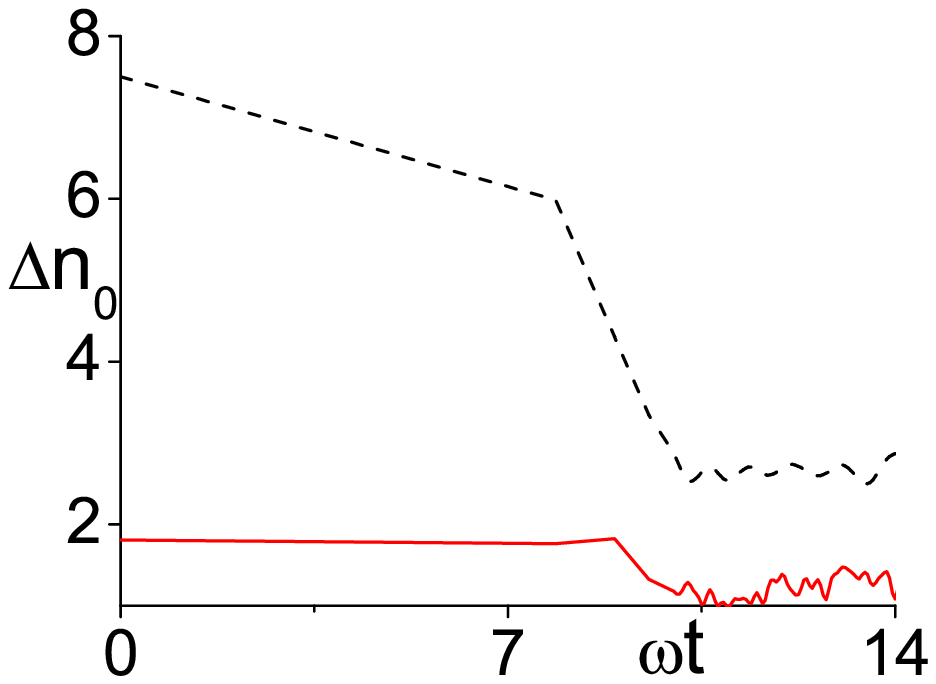}
\includegraphics[width=0.49\columnwidth]{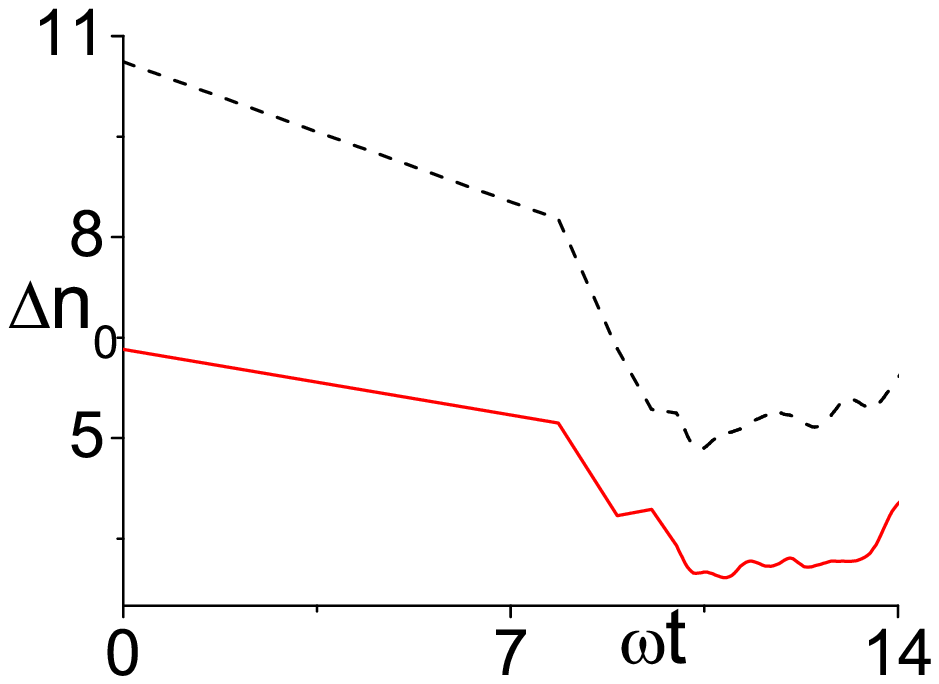}
\vspace{-0.4cm} \caption{The phase coherence and the atom number
fluctuations, obtained by taking into account the symmetric operator
ordering in the TWA (dashed line) and by assuming the stochastic
representation to be normally ordered (solid line) for an initial
temperature of the system $k_BT_i/\hbar\omega=6.67$ (left) and
$38.46$ (right). The same system as in
Fig.~\ref{sampling}.}\label{P-functionfig}
\end{figure}

We illustrate with a numerical example the importance of the
particular stochastic representation of the density matrix. We
integrate both TWA and TPA according to Eq. \eqref{GPE} for the same
system at finite initial temperature $T_i$ by means sampling the
Wigner and the $P$ distributions. The results shown in
Fig.~\ref{P-functionfig} for the phase coherence $C_1$ and the atom
number fluctuations $\Delta n_0$ are then useful in understanding
the importance of the operator ordering. The P-distribution does not
introduce the correct noise in the initial state and therefore it
fails to produce the right phase coherence properties and
underestimates the atom number fluctuations. At high temperature,
the TPA is closer to the TWA because thermal noise starts dominating
vacuum fluctuations.

\section{Concluding Remarks}\label{remarks}

We studied the loading of a harmonically trapped BEC into an optical
lattice. As the lattice height is increased, quantum fluctuations
become more important and the classical GP description breaks down.
We found that the TWA provides a powerful tool to study
nonequilibrium dynamics of bosonic atoms in an optical lattice. We
showed that, especially at low temperatures, the correct quantum
statistics of phonon modes and the accurate treatment of the
operator orderings in the Wigner representation are very important.

The TWA incorporates the full multi-band description of the lattice
dynamics and becomes more accurate when the occupation number of the
sites is large and quantum fluctuations are not too dominating. In
the TWA simulations of the turning up of the lattice potential we
found the atom number squeezing to saturate for deep lattices, which
can be explained by the finite turning-up time. It is numerically
more demanding to study significantly longer ramping-up times that
would result in considerably more reduced atom number fluctuations.
It is also expected that the TWA would eventually break down over
longer time scales very close to the MI state that could be
interesting to investigate. In our theoretical study of dissipative
dipole oscillations of bosonic atoms \cite{RUO05}, we successfully
modeled the system with the TWA for the case of considerably
stronger quantum fluctuations than in the present work, in a good
agreement with the experimental results \cite{FER05}. This seems to
suggest that also notably stronger atom number squeezing could be
studied within the TWA than the one reported in this paper by
considering very long ramping-up times. As our analysis shows,
however, that, in the case of lattices with large filling factors,
very long ramping-up times, that are required to reach the MI ground
state, can be extremely challenging in actual experiments.

The TWA simulations could also be extended to 2D or 3D lattices. In
2D and 3D the implementation of the TWA may require a special care,
but, at least in the uniform space, it has been successful in
modeling thermal fluctuations \cite{SIN02}. In higher dimensions
also more complex initial states could be considered, such as
topological defects and textures that may be prepared by phase
imprinting \cite{RUO01}.

\appendix

\section{Bogoliubov theory in a discrete tight-binding approximation}\label{bogo-approach}

It is useful to compare the TWA simulation results to the ground
state fluctuations of the atoms, obtained from the Bogoliubov
theory. In a deep optical lattice, with very large $s$, and close to
the ground state only one mode per lattice site is important and the
system can be approximated by the discrete Bose-Hubbard Hamiltonian
(BHH), where we expand the field operator on the basis of the
Wannier functions and keep only the lowest vibrational states in
each lattice site, $\h\psi(x)=\s_i\h b_i\eta_i(x)$:
\beq
H=\s_i\big[\nu_i \h b^\d_i\h b_i-J(\h b^\d_i\h b_{i+1}+ {\rm
H.c.})+\f{U}{2}(\h b_i^\d)^2\h b_i^2\big]\,, \label{BH}
\eeq
where the summation is over the lattice sites, $J\simeq -\int dx
\eta^*_i(x){\cal L}\eta_{i+1}(x)$ is the hopping amplitude between
the nearest-neighbor sites, $U\simeq g_{1D}\int dx |\eta_j(x)|^4$ is
the on-site interaction constant, and $\nu_j\equiv j^2 d^2
m\omega^2/2$ represents the harmonic trapping potential, with $j=0$
site at the trap center. We may approximate the Wannier functions
$\eta_i$ by the ground state harmonic oscillator wave function with
the frequency
\beq
\label{omega s} \omega_s=\f{2s^{1/2}E_r}{\hbar}\,,
\eeq
which is obtained by expanding the optical potential at the lattice
site minimum \cite{JAK98}. Analytic approximations for $U$ and $J$
may be derived using the Gaussian approximations to the Wannier wave
functions or the Matthieu functions of the energy band. When we
compare the TWA results to the BHH, we frequently extract the
expectation values involving $\h{b}$ using Eq.~(\ref{projection})
with $\h{b}\sim\h{a}$. We tested that using different projections
does not affect the results. For $n_i J\agt U$, with $n_i\equiv\< \h
b^\d_i \h b_i\>$, the system is in the superfluid regime with the
long-range phase coherence and is expected to undergo the MI
transition at $2.2 n_i J\simeq U$ \cite{FIS89,ZWE03}, resulting in a
highly number squeezed ground state.

In the Bogoliubov approach to the BHH we introduce the discrete
position coordinate is $x=jd$, where $j$ is the integer labeling the
lattice sites. We decompose the ground state and the excited state
fractions into mutually orthogonal subspaces and study the
linearized fluctuations around $c_j\equiv\< \h b_j\>$:
\begin{align}
\h b_j &= c_j+\delta \h b_j \nonumber \\ &=c_j+  \sum_n [f_n(jd)
\h\chi_n-h_{n}^*(jd)\h\chi_n^\d]\,,\label{bhhbogo}
\end{align}
where $\chi_n$ are the quasiparticle operators and $[f_n(jd),$
$h_n(jd)]$ are the Bogoliubov modes, obtained from:
\begin{align}\label{Bogolat}
\mathcal{L}^{d}_j f_n^j+J(f_n^{j+1}+f_n^{j-1})-c_j^2Uh_n^j   &
= \epsilon_j f_n^j \nonumber \\
\mathcal{L}^{d}_j h_n^j+J(h_n^{j+1}+h_n^{j-1})-c_j^{*2}Uf_n^j & =
-\epsilon_j h_n^j \, ,
\end{align}
where $\mathcal{L}^{d}_j \equiv 2n_jU+\nu_j-\mu$ and $f_n^j\equiv
f_n(jd)$.

The expression for the fluctuations $\delta\h b_j$ in \eq{bhhbogo}
may be inverted by using the relation:
\beq
\sum_j  \[ f_n(jd)f_{m}^*(jd)-h_n(jd)h_m^*(jd) \] =\delta_{mn} \,,
\eeq
in order to obtain $\h \chi_n$:
\beq
\h \chi_n=\sum_j \[ f_n(jd)\delta\h b_j + h_n(jd)\delta\h
b_j^\d\]\,.\label{chiexpr}
\eeq
In Section \ref{ret-lat} we use this expression to evaluate the
occupation numbers
\beq
\overline n_k=\< \h \chi_k^\d\h \chi_k\>
\eeq
of the low energy quasiparticle modes in the lattice in the TWA
simulations by substituting $\delta \h b_j= \h b_j - c_j$, so that
$\overline n_k$ is obtained from the numerically evaluated
correlation functions involving $\h b_j$'s.

The atom number operator $\h n_i$ at the site $i$ may be obtained by
noting that we may expand to first order in $\delta \h b_i$:  $\h
b_i^\d \h b_i\simeq n_i+\h n_i$, for
\beq
\h n_i= \sqrt{n_i} \sum_j (w_j \h\chi_j+ w^*_j\h\chi^\d_j)\,,
\eeq
where $w_j\equiv f_j-h_j$. Moreover, we introduce the phase operator
at the site $i$ as
\beq
\h \varphi_i= -\f{i}{2 \sqrt{n_i}} \sum_j (r_j\h\chi_j -
r_j^*\h\chi^\d_j)\,,
\eeq
for which the commutator $[\h n_j, \h \varphi_j]=i$ and we have
defined $r_j\equiv f_j+h_j$.

Diagonalizing Eqs.~\eqref{Bogolat} yields the results for the number
fluctuations in the $i$th site and the phase fluctuations between
the $k$th and $l$th site:
\beq
\label{deltanbogo} (\Delta n_i)^2= n_i \sum_j |w_j|^2
(2\bar{n}_j+1)\,,
\eeq
and
\begin{align}\label{deltaphibogo}
(\Delta\varphi_{kl})^2 &\equiv
\<(\h\varphi_k-\h\varphi_l)^2\>\nonumber
\\ &= \f{1}{4}\sum_j \Bigg |\f{r_j(kd)}{\sqrt{n_k}}-\f{r_j(ld)}{\sqrt{n_l}}\Bigg |^2
(2\bar{n}_j+1)\,,
\end{align}
where $\bar{n}_j=\<\h\chi^\d_j\h\chi_j\>$ is the thermal population
of the $j$th phonon mode in the lattice.

In an inhomogeneous BEC, the phonon modes are spatially localized
and Eqs.~\eqref{deltanbogo} and~\eqref{deltaphibogo} generally lead
to spatially varying density and phase fluctuations. However, in the
homogeneous case ($\nu_i=0$) the Bogoliubov expressions for the
phase and number fluctuations are suitable for an analytic
treatment. In the case of the homogeneous system with $n$ atoms per
site, we obtain the excitation frequencies~\cite{JAV99,BUR02}:
\beq\label{homolat}
(\hbar\omega_q)^2=4J\sin^2{\(\f{qd}{2}\)}\[4J\sin^2{\(\f{qd}{2}\)}+2nU\]\,,
\eeq
where $q$ is the quasiparticle momentum with periodic boundary
conditions, so that $qd= 2\pi m/N_p$, for $m=-N_p/2,\ldots,
N_p/2-1$, and $N_p$ denotes the number of lattice sites. Moreover,
in the experimentally interesting regime $nU\gg J$ we obtain
\beq
\hbar\omega_q\simeq 2\sqrt{2nJU} \left|\sin{\(qd/2 \)}\right|\,,
\eeq
and for $N_p\gg1$ the lowest phonon mode energy
\beq
\hbar \omega_{q,{\rm min}}\simeq {2\pi \sqrt{2nJU}\over
N_p}\,.\label{lowestphonon}
\eeq
Moreover, in the limit $nU\gg J$, the expressions for the number and
phase fluctuations in Eqs.\ \eqref{deltanbogo}
and~\eqref{deltaphibogo} can be approximated by \cite{JAV99,BUR02}:
\begin{align}
\label{deltan} (\Delta n_i)^2\simeq &\sum_q \f{\hbar\omega_q}{2UN_p}
(2\bar{n}_q+1),\\ \label{deltaphi} (\Delta \varphi_{i,i+1})^2\simeq
& \sum_q \f{\hbar\omega_q}{4nJN_p} (2\bar{n}_q+1)\,,
\end{align}
where $\bar n_q$ is the occupation number of the phonon mode with
the quasimomentum $q$. By replacing the sum by an integral when
$N_p$ is large, these yield at $T=0$:
\begin{align}
\label{deltan-T0} (\Delta n_i)^2\simeq & \f{1}{\pi}
\(\f{8nJ}{U}\)^{1/2},\\
\label{deltaphi-T0} (\Delta \varphi_{i,i+1})^2\simeq &
\f{1}{2}\(\f{2U}{nJ}\)^{1/2}\,.
\end{align}
Formulas \eqref{deltan-T0} and \eqref{deltaphi-T0} provide useful
comparisons to our numerical results.
\begin{figure}
\includegraphics[width=0.49\columnwidth]{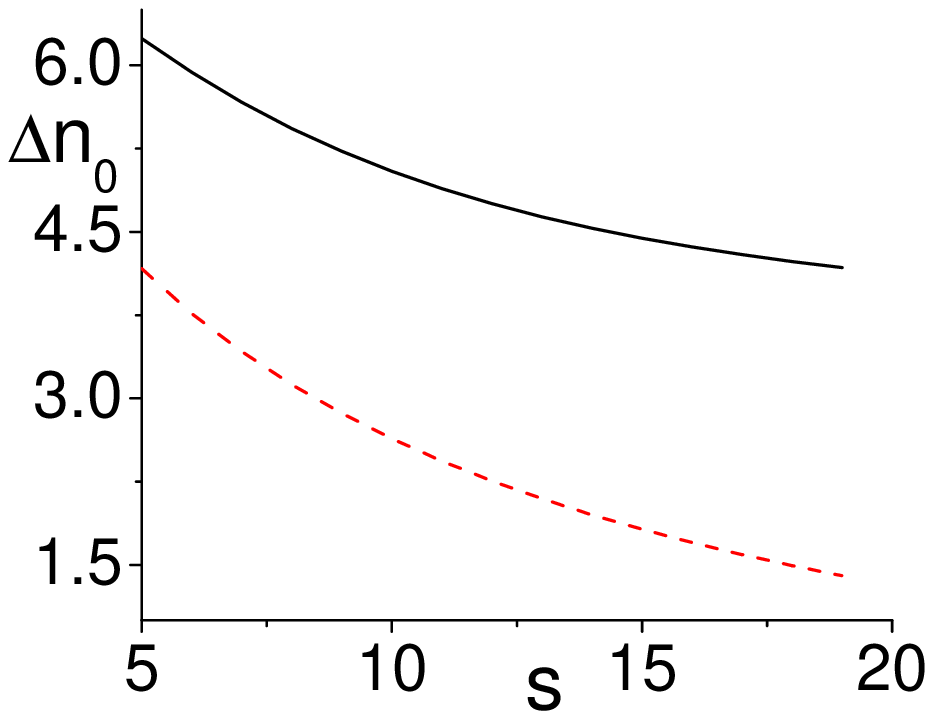}
\includegraphics[width=0.49\columnwidth]{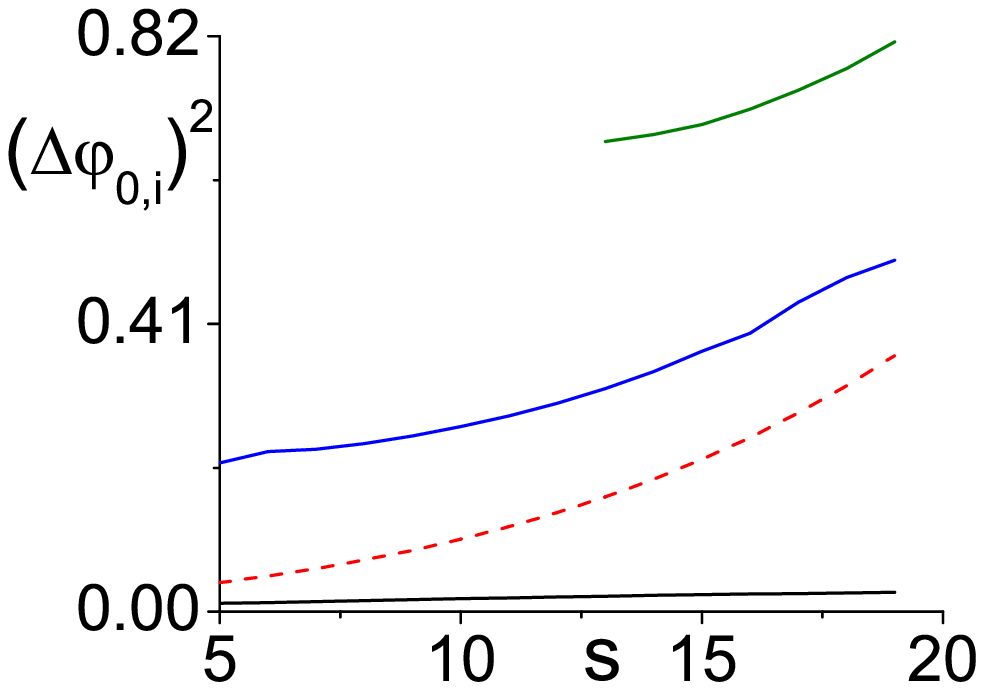}
\vspace{-0.4cm}\caption{The number and phase fluctuations of atoms
in a combined harmonic trap and optical lattice, obtained by
numerically solving the Bogoliubov modes in the discrete
tight-binding approximation. On left we show $\Delta n_0$ for
various values of $s$ (solid line) and the analytic homogenous
result with $n=n_0$ in Eq.~\eqref{deltan-T0} (dashed line). Here the
total atom number $N_0=2000$, the nonlinear constant
$g_{1D}=0.05\hbar\omega l$, $T=0$, and $n_0\simeq$80-90 atoms. On
right for the same inhomogeneous system we show approximate
$(\Delta\varphi_{0,i})^2$ for $i=1,13,16$ (solid lines from bottom)
and the analytic homogeneous result for $(\Delta\varphi_{0,1})^2$
(dashed line) from Eq.~\eqref{deltaphi-T0}. The inhomogeneous
results for $(\Delta\varphi_{0,i})^2$ should only be understood
qualitatively as an order of magnitude
estimates.}\label{num-bogolat}
\end{figure}

In Fig.~\ref{num-bogolat} we show the numerical results of the
number and phase fluctuations in a harmonic trap, obtained by
solving the Bogoliubov approximation to the BHH from
Eqs.~\eqref{Bogolat} for a typical system in our simulations. We
evaluate $\Delta n_0$ using Eq.~\eqref{deltanbogo} and compare it to
the homogeneous results with $n=n_0$, determined by
Eq.~\eqref{deltan-T0}. We also calculate $(\Delta\varphi_{0,i})^2$
as a function of the lattice height $s$ using
Eq.~\eqref{deltaphibogo} for several sites along the lattice and we
evaluate $(\Delta\varphi_{0,1})^2$ in the homogeneous case from
Eq.~\eqref{deltaphi-T0}. The inhomogeneous results for the phase
fluctuations are numerically very unstable due to the summation
formula in \eq{deltaphibogo} and spatially localized phonon modes.
Consequently, the results for $(\Delta\varphi_{0,i})^2$ should only
be considered as an order of magnitude estimates, describing the
qualitative behavior. Numerically we find the Bogoliubov result in a
harmonic trap for $\Delta n_0$ to be larger and for
$\Delta\varphi_{01}$ smaller than the homogeneous result. The
harmonic trap significantly reduces phase fluctuations close to the
trap center and enhance them close to the edge of the atom cloud.

\section{Three-body losses}\label{3body}

The TWA approach in this paper ignored any atom losses due to
three-body collisions. The TWA with incorporated three-body losses
would result in a dynamical stochastic noise term in \eq{GPE} and
more demanding numerics. We may estimate the importance of the
three-body losses in an optical lattice system as in
Ref.~\cite{JAC03}. In a deep lattice with a fragmented condensate
the atom loss rate at the central lattice site (with $n_0\gg 1$) is
approximately
\beq
{dn_0\over dt}\simeq -\Gamma n_0^3,\quad \Gamma={K_3\over 12} \int
d^3r\,|\eta_0(\rv)|^6\,, \label{losses}
\eeq
where $K_3$ is the recombination event rate and $\eta_0(\rv)$ is the
ground state wave function in the central site with the atom number
$n_0$. For simplicity, we have ignored the correlations between
different lattice sites.

We obtain from \eq{losses}:
\beq
\Gamma\simeq {K_3\sqrt{s}\over 144\sqrt{3}\pi a^2 d^2 l^2}\(
{g_{1D}\over \hbar\omega l}\)^2\simeq {67 K_3\over a^2 l^4}\,.
\eeq
Here on the right-hand side we have introduced the same parameters
as in Fig.~\ref{heightfig} with $s=20$. In order to the three-body
loss rate to be negligible over the time scale of the simulations we
require $\Gamma n_0^3\ll \omega$ that may be satisfied for
sufficiently large $l$. For the 1D description of the dynamics to be
valid we also need to maintain $\hbar\omega_\perp \agt n U\sim n
g_{1D}ks^{1/4}/\sqrt{2\pi}$. For example, for $^{87}$Rb we use
$K_3\simeq 2.2\times10^{-28}$ cm$^6$/s \cite{SOD99} and
$a\simeq5.313$nm. Then $\omega=2\pi\times1$Hz yields $\Gamma
n_0^3\simeq 0.006 \omega$ and, using an average occupation number
$n$, $\hbar\omega_\perp \sim 3 n g_{1D}ks^{1/4}/\sqrt{2\pi}$ with
$\omega_\perp\gg \omega$. The effect of the three-body losses could
be further reduced by using different atoms and the Feshbach
resonances.

\acknowledgments{We acknowledge financial support from the EPSRC.}


\begin{thebibliography}{99}
\bibitem{rethcom} In the studied 1D system we never
observed any considerable rethermalization on the time scale of our
simulations. This is quite different from the TWA simulations in 3D
\cite{SIN02} and may be the consequence of the smaller density of
states in 1D.
\end{thebibliography}
\end{document}